\documentclass{cernrep}
\usepackage[colorlinks=true, linkcolor=black, citecolor=black, urlcolor=blue]{hyperref}
\usepackage{fancyhdr}
\fancyhfoffset{4 mm}
\fancypagestyle{ARTTITLE}{%
\fancyhf{} 
\lhead{\footnotesize{Proceedings of the 2019 CERN--Accelerator--School course on
\it{ High Gradient Wakefield Accelerators}, Sesimbra, (Portugal)}}
\lfoot{Available online at \url{https://cas.web.cern.ch/previous-schools}}
\rfoot{\thepage\hspace*{3mm}}

   }
\begin{document}

\title{Introduction to Particle Accelerators and their Limitations}

\author{Massimo Ferrario$^1$, Bernhard J. Holzer$^2$}
\institute{$^1$Istituto Nazionale di Fisica Nucleare - Laboratori Nazionali di Frascati, Rome, Italy; $^2$CERN, Geneva, Switzerland}
\begin{abstract}
The paper gives a short overview of the principles of particle accelerators, their historical development and the typical performance limitations. After an~introduction to the basic concepts, the main emphasis is to sketch the layout of modern storage rings and their limitations in energy and machine performance. Examples of existing machines - among them clearly the LHC at CERN - demonstrate the basic principles and the~technical and physical limits that we face today in the design and operation of these particle colliders. Pushing for ever higher beam energies motivates the design of the future collider studies and beyond that, the~development of more efficient acceleration techniques.
\end{abstract}
\maketitle
\thispagestyle{ARTTITLE}
\section {Introduction} 

The study of matter, starting from the structure of the atom, the discovery of the nucleus and beyond that the discovery of the variety  of elementary particles and their interactions, summarised in a scientific picture that we like to call "Standard Model", came along and drove the development of powerful tools to create high energetic particle beams, the so-called particle accelerators.
\par
These machines are used nowadays world wide in engineering, life science and physics.
More than 35000 accelerators exist today for studies in a manifold of applications in physics, chemistry, medicine and structure analysis. 
The up to now largest accelerator, the Large Hadron Collider, LHC, at CERN in Geneva is operating on a level of 7 TeV energy per beam, corresponding to an available centre of mass energy of $E_{cm}=14$ TeV.
It is standing in a long tradition of technical as well as physics progress in the~creation of high energy particle beams, their acceleration and the~successful collision of the micro meter small beam sizes. 
This article gives a very basic introduction into the physics of these accelerators and (some of) their limitations.
They have been selected arbitrarily by the authors, de-facto there are many more that can be found and studied by the reader in a number of decent publications ( $\ldots$ and somewhen hopefully overcome).
We define a few key parameters to describe the performance of an~accelerator. They clearly depend on the~application, however, on a quite general basis we can summarize them as: 
\begin{itemize} 
\item energy of the accelerated particles
\item particle intensity, or current of the stored or accelerated particle beam
\item beam quality, the experts call it {\it emittance} 
\item in the case of lepton beams, used for the production of synchrotron radiation, brightness of the~emitted light
\item luminosity in case of particle colliders.
\end{itemize}

\begin{figure}[hb]
\centering\includegraphics[width=0.60\columnwidth]{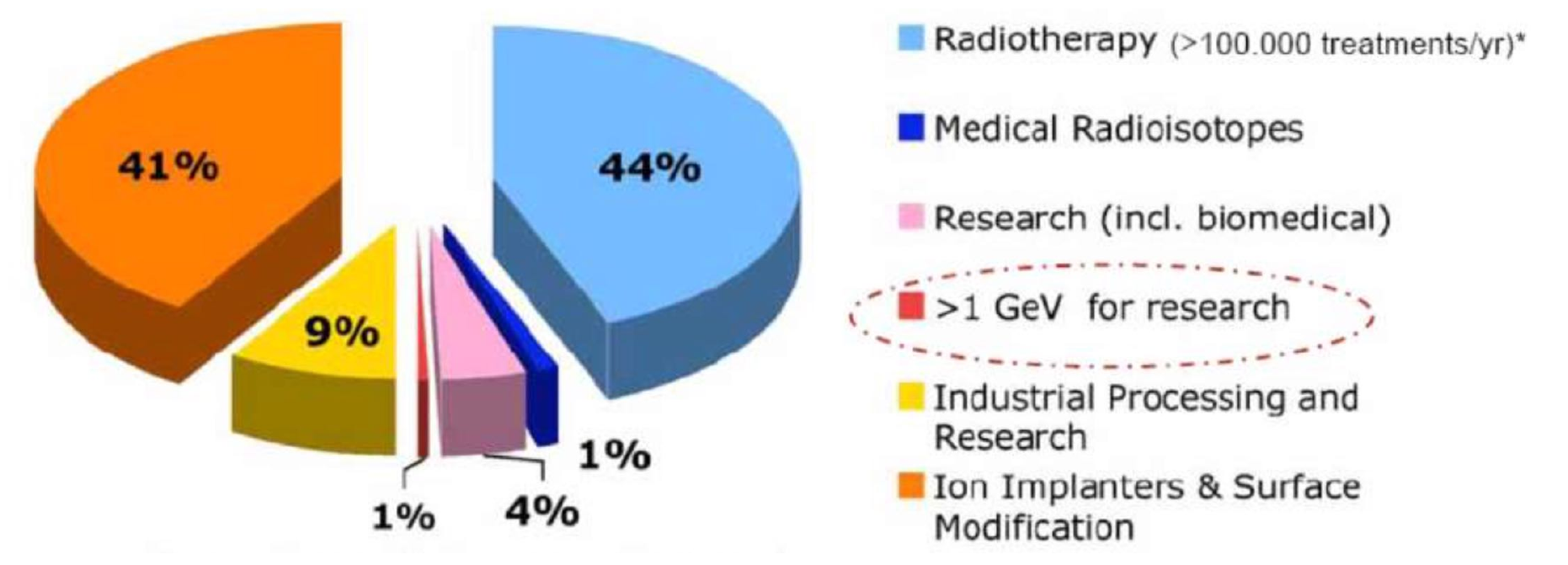}
\caption{Particle accelerators and their applications.}
\label{applications}
\end{figure}

\begin{figure}[h!]
\begin{center}
\includegraphics[width=0.70\columnwidth]{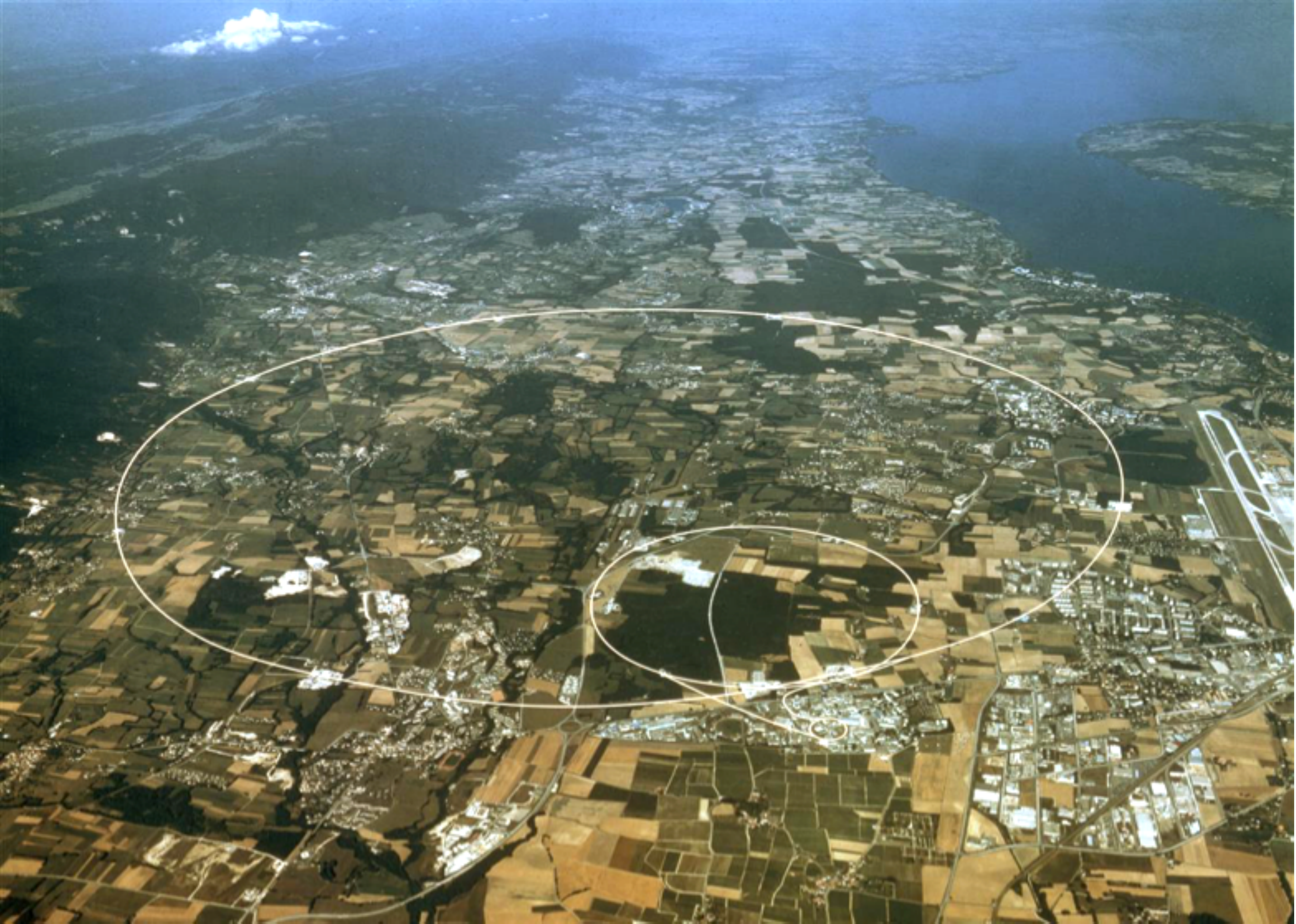}
\caption{The Large Hadron Collider, LHC, located in the Geneva valley.}
\label{lhc_geo}
\end{center}
\end{figure}

As a matter of fact, it depends on the application of the particle beam, which of these parameters are the (most) relevant ones and how we have to optimise the design of our accelerator to reach highest performance. Out of the 35000 particle accelerators that exist and operate worldwide, the~largest fraction is used for medical applications and ion implantation. Figure~\ref{applications} shows the~distribution of the existing machines over their field of operation. 
We must admit, that for those, working in basic research, like e.g., the authors of this article, it is sometimes brings us down to earth to realise that the number of machines working at the energy frontier is "close to negligible". Still, and be it for the bare impression of such a~large piece of scientific infrastructure, we show in Fig.~\ref{lhc_geo} a photo of the LHC in the Geneva valley.

In this paper we would like to follow a little bit the path that has been paved by the ingenious discoveries and developments that date back to the time of the discovery of the nucleus by Ernest Rutherford.

Using alpha particles of some MeV from a natural source, he discovered in scattering experiments the nucleus of the atom.  Now, a radioactive source is not an ideal tool for precise, trigger-able and healthy experiments. So it was indeed Rutherford who discussed the possibility of artificially accelerated particles with two colleagues, Cockcroft and Walton. Following this idea, they developed within only four years the first particle accelerator ever built: In 1932 they could demonstrated the first splitting of a~nucleus (Li) using a 400 keV proton beam. 
\par 
Their acceleration mechanism was based on a rectifier- or Greinacher-circuit, consisting of a number of diodes and capacitors that transformed a relatively small AC-voltage to a DC potential that corresponds to a multiple of the applied basic potential, according to the number of diode/capacitor units that are used.
The particle source was a standard hydrogen discharge source, connected to the high-voltage part of the system and the particle beam was accelerated to ground potential, hitting the Lithium target \cite{Cockcroft}. Even if of modest energy at that time, their beam could split the Li nucleus and \dots 
needles to say: Cockcroft and Walton got the Nobel-prize for it.

\subsection {Limit I: Beam Energy: Voltage Breakdown in DC Accelerators}
%

\begin{figure}[h!]
\begin{center}
\includegraphics[width=0.4\columnwidth]{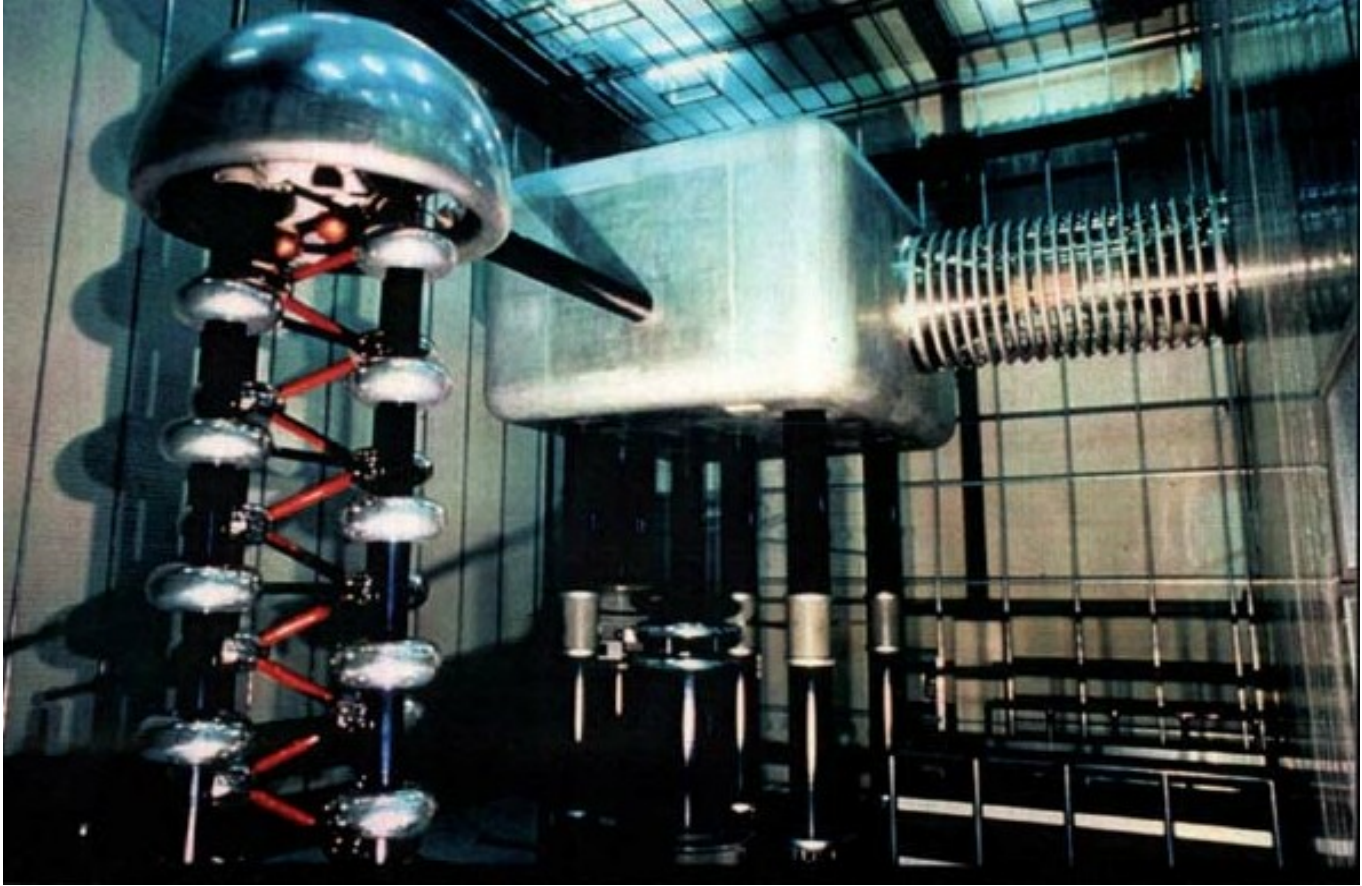}
\caption{A Cockcroft-Walton Generator used at CERN as pre-accelerator for the proton beams. The device is meanwhile replaced by the more compact and efficient Radio Frequency Quadrupole (RFQ) technique.
}
 \label{Cockcroft_photo}
\end{center}
\end{figure}

\begin{figure}[h!]
\begin{center}
\includegraphics[width=0.7\columnwidth]{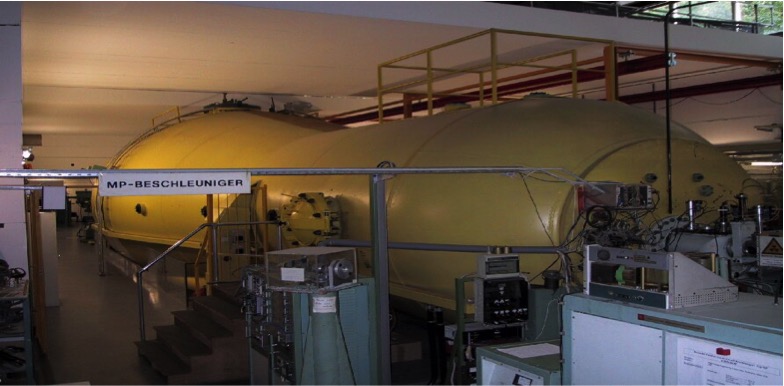}
\caption{A typical example of a Tandem-van-de-Graff Accelerator. These are very reliable machines for precision measurements in atom and nuclear physics. (Photo: Max Planck Institute for Nuclear Physics, Heidelberg)%
}
 \label{vdg_photo}
\end{center}
\end{figure}

In a parallel approach, but based on a completely different technique, another type of DC accelerator had been invented: Van-de-Graaf designed a DC-accelerator \cite{vdGraaf}, that used a mechanical transport system to carry charges, sprayed on a belt or chain, to a high voltage terminal. In general these machines reach higher voltages than the Cockcroft-Walton devices but they are more limited in particle intensity. Common to all DC accelerators is the limitation of the achievable beam energy due to high voltage breakdown effects (discharges). Without insulating gas ($SF_6$ in most cases) electrical fields will be limited to about 1 MV/m and even using most sophisticated devices, like the one in  Fig. \ref{vdg_photo}, acceleration voltages of some MV cannot be overcome. 


In fact, the example in the last figure shows an option that is applied in a number of places: Injecting a negative ion beam (even $H^{-}$ is used) and stripping the ions in the middle of the high voltage terminal allows to profit from the potential difference twice and thus to gain another step in beam energy.

\subsection{Limit II: The size of the accelerating structure}
\begin{figure}[h!]
\begin{center}
\includegraphics[width=0.7\columnwidth]{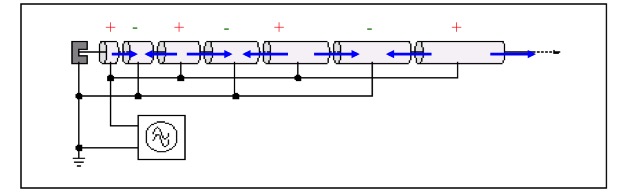}
\caption{Schematic view of the Wiederoe principle as fundamental concept for AC (or "RF") acceleration.%
}
\label{Wiederoe_schema}
\end{center}
\end{figure}

Given the obvious limitations of the above described DC machines, the next step forward is a~natural choice: \\
In 1928 Wideroe developed the concept of an AC accelerator. Instead of rectifying the AC voltage, he connected a series of acceleration electrodes in alternating order to the output of an AC supply. The~schematic layout is shown in  Fig. \ref{Wiederoe_schema} where - for a moment the direction of the electric field is indicated by the arrows. 

In principle this device can produce step by step a multiple of the acceleration voltage, as long as for the negative half wave of the AC voltage the particles are shielded from the decelerating field. The~energy gain after the $n^{th}$ step therefore is
\begin{equation}
E_n=n\cdot q \cdot U_{0} \cdot sin(\psi_s),
\end{equation}
where n denotes the $n^{th}$ acceleration step, q the charge of the accelerated particle, $U_{0}$ the applied voltage per gap and $\psi_s$ the phase between the particle and the changing AC-voltage, as indicated in  Fig. \ref{Wiederoe_half_wave}

The key-point in such a Wideroe-structure is the length of the drift tubes that will protect the particles from the negative half wave of the sinusoidal AC voltage. For a given frequency of the~applied RF voltage the length of the drift tube is defined by the speed of the particle and the~duration of the~negative half wave of the sinusoidal voltage: 

\begin{figure}[h!]
\begin{center}
\includegraphics[width=0.45\columnwidth]{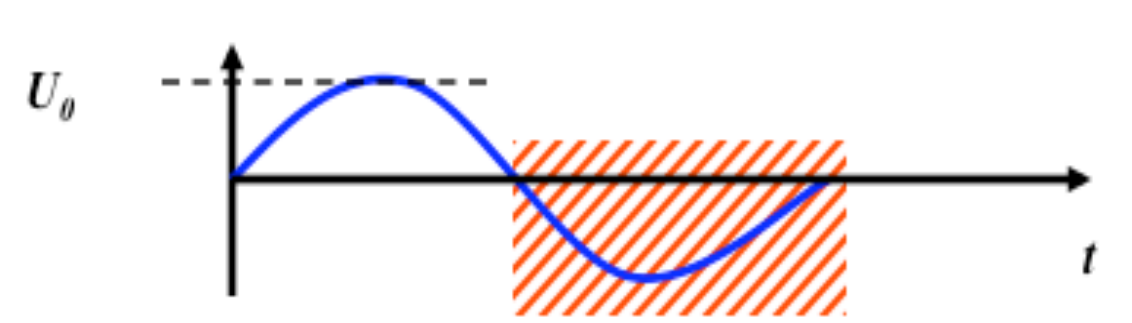}
\caption{The frequency, and so the period of the RF system, and the particle speed determine the length of the~drift tubes in the structure.
}
\label{Wiederoe_half_wave}
\end{center}
\end{figure}
The time span of the negative half wave is defined by the applied frequency, $ \Delta t=\tau_{rf}/2$ , where 
$\tau_{rf}$ is the RF period. And so we get for the length of the $n^{th}$ drift tube

\begin{equation}
l_{n}=v_{n} \cdot \frac{\tau_{rf}}{2}.
\end{equation}

It depends only on the period of the RF system and the velocity of the particle $v_n$ when traversing the $n^{th}$ acceleration gap.  Given the kinetic energy of the particle of mass $m$ and velocity~$v$,

\begin{equation}
E_{kin}=\frac {1}{2} \cdot mv^{2}
\end{equation}

we obtain directly 
\begin{equation}
l_{n}=\frac {1} {\nu_{rf}} \cdot \sqrt{\frac {nqU_{0}sin{\psi_{s}}}{2m}},
\end{equation}
which defines the design concept of the machine.

\begin{figure}[h!]
\begin{center}
\includegraphics[width=0.5\columnwidth]{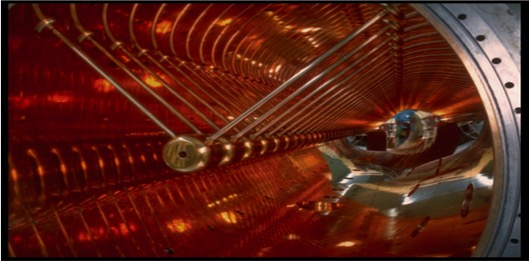}
\caption{Unilac at GSI, Darmstadt; the structure of the drift tubes and their increasing length as a function of the~particle energy is clearly visible. %
}
\label{Unilac}
\end{center}
\end{figure}

  {Figure \ref{Unilac}} shows a photograph of such a device, the Unilac at the GSI institute in Darmstadt.

Two remarks should be made in this context:
\begin{itemize}
\item This short derivation here is based on the classical approach. And in fact: These accelerators are usually optimum for "low energy" proton or heavy ion beams. Typical beam energies - referring to protons -  are in the range of some 10 MeV. E.g. the  Linac 2 at CERN delivers the protons for LHC operation with an energy of 50 MeV, corresponding to a relativistic beta of $\beta = 0.31 $. 
\item For higher energies, even in the case of protons or ions the speed some-when will approach the~speed of light and the length of the drift tubes and so the dimension of the whole accelerator reaches considerable sizes that might not be feasible anymore. A more advanced concept is needed in order to keep the machine within reasonable dimensions. And the next logic step in the~historical development was to introduce magnetic fields to bend the trajectory of the particle beam onto a circle.    
\end{itemize}

\section{Beam Dynamics in Synchrotrons and Storage Rings}
\subsection{The quest for highest energies}
Before we continue in our rush for higher and higher energies, it seems adequate to motivate this never ending effort by two trivial (... just another word for "well known") statements:
A high energetic particle beam allows us to study new manifestations of matter. 
Having a certain amount of energy $E$ available allows to create a given amount of mass $m$:
\begin{equation}
E=mc^2,
\end{equation}
where $c$ is the speed of light. As easy as that. And the quest for understanding the underlying structure of matter, means the creation and study of its fundamental constituents or building blocks: the mesons, baryons, and finally the quarks and leptons.

There is a second motivation to achieve highest possible beam energies that is somehow de-coupled from the bare need to create new particles: The resolution that we can achieve in particle scattering experiments. Quantum mechanics tells us that a high energetic particle can and has to be considered as a particle-wave with a wavelength defined by the momentum of the particle. The~so-called de-Broglie wave length is given by 
\begin{equation}
\lambda = \frac {h}{p},
\end{equation}
where { \it h} is the Planck constant,  $h =6.626*10^{-34}$ Js and {\it p} the particle momentum.
For example, and to put the things in the right perspective: the famous green light of the visible spectrum corresponds to a wavelength of about $\lambda =500$ nm.
On the other side, the SLAC linear accelerator in Stanford, pushed the electrons to an energy of 50 GeV and on this level their de-Broglie wavelength corresponds to  $\lambda =1.4 * 10^{-17}$ m.
Now, while atoms have a typical radius of a~few~$\AA  = 10^{-10}$~m, nuclei are in the order of  $10^{-15}$ m. It is no surprise that Kendal, Friedman and Taylor could discover the quarks using the SLAC electron beam, but they never would have had a chance to get their Nobel prize  using a~conventional microscope.
\par
Having made this point, we can  
consider the next step forward in achievable beam energy, which requires circular structures. In order to apply over and over again the accelerating fields we will try to bend the particles on a circle and so bring them back to the RF structure where they receive the next step in energy.
\par
As a consequence we will have to introduce magnetic ($\vec B$) or electric ($\vec E$) fields that deflect the~particles and keep them during the complete acceleration process on a well defined orbit. The~Lorentz-force that acts on the particles thus will have to compensate exactly the centrifugal force that the particles will feel on their bent orbit.
\\
In general we can write for a particle of charge $q$: 
\begin{equation}
\vec F = q \cdot (\vec E+ \vec v \times \vec B).
\end{equation}
Now, talking about high energy particle beams the velocity $\it{v}$ is close to the speed of light and so represents a nice amplification factor whenever we apply a magnetic field. As a consequence magnetic fields for bending and focusing of high energy charged particles are much more convenient than electric fields.
\\
Neglecting electrical fields, therefore, at the moment we write the Lorentz force and the centrifugal force of the particle on its circular path as 
\begin{equation}
F_{Lorentz}=e \cdot v \cdot B  
\end{equation}
\begin{equation}
F_{centrifugal}=\frac{\gamma m_{0} v^{2}}{\rho},
\end{equation}
where $m_0$ stands for the particle's rest mass, $\rho$ describes the bending radius of the trajectory and $\gamma$ is the relativistic Lorentz factor. Assuming an idealised homogeneous dipole filed along the particle orbit we define the condition for a perfect circular orbit as equality between these two forces to obtain the~condition for the idealised ring:
\begin{equation}
\frac{p}{e}={B \cdot \rho}, 
\end{equation}
where we refer to protons and accordingly have set {\it q = e}.
This condition relates the so-called beam rigidity $B\rho$ to the particle momentum that can be carried in the storage ring, and it will in the end define -- for a given magnetic field of the dipoles -- the size of the storage ring.

\begin{figure}[h!]
\begin{center}
\includegraphics[width=0.35\columnwidth] {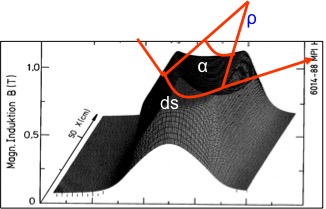}
\caption{Dipole field of a storage ring and the schematic path of the particles.%
}
\label{TSR_dipole_field}
\end{center}
\end{figure}

Now in reality instead of a continuous dipole field the storage ring will be built out of several dipoles, powered in series to define the geometry of the ring.
For a single dipole magnet the particle trajectory is shown schematically  in   Fig. \ref{TSR_dipole_field}.

In the free space outside the dipole magnet the particle trajectory is following a straight line. As soon as the particle enters the magnet it is bent on a circular path until it leaves the magnet at the other side.

\subsection{Limit III: The Magnetic Guide Field}
The overall effect of the main bending (or "dipole") magnets in the ring will define a more of less circular path that we call design orbit. By definition this design orbit has to be closed upon itself and thus the~overall effect of the main dipole magnets in the ring has to define a bending angle of exactly $2 \pi$. 
If $\alpha$ defines the bending angle of one single magnet 
\begin{equation}
\alpha=\frac{ds}{\rho}=\frac{Bds}{B \cdot \rho}
\end{equation}
we require therefore 
\begin{equation}
\frac{\int B dl}{p/e} = \frac{\int B dl}{B \cdot \rho}=2 \pi.
\label{eq_beam_rigidity}
\end{equation}
In the true sense of the word a storage ring, therefore, is not a ring but more a polygon, where "poly" stands for the number of dipole magnets installed in the "ring". 

\par
\begin{figure}[h!]
\begin{center}
\includegraphics[width=0.35\columnwidth] {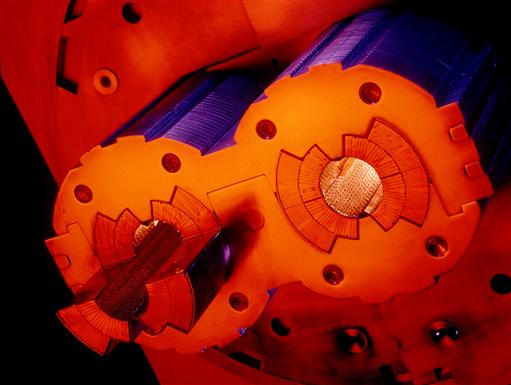}
\caption{Superconducting dipole magnets of the LHC.%
}
\label{LHC_dipole}
\end{center}
\end{figure}

 
In the case of the LHC the dipole field has been pushed to the highest achievable values. 
1232 super conducting dipole magnets, with a length of 15 m each, define the geometry of the ring and via Eq. (\ref{eq_beam_rigidity})  the maximum momentum for the stored proton beam.
Following the equation given above we obtain for a maximum momentum of $p=7\  TeV$ a required B-field of 
\begin{equation}
B=\frac {2 \pi \cdot 7000 10^{9} eV}{1232 \cdot 15 \ m \cdot 2.99792 \cdot 10^{8} \ m/s}
\end{equation}

\begin{equation}
B=8.33 \ T
\end{equation}
that is needed to bend the beams. 
For convenience we have expressed the particle momentum in units of GeV/c.
 {Figure \ref{LHC_dipole}} shows a photo of the LHC dipole magnets, built out of super conducting NbTi filaments, that are operated at a temperature of T=1.9 K. 

It is clear from Eq. (\ref{eq_beam_rigidity}) that in case we are limited to a certain maximum achievable field of the~bending magnets, the only way to accelerate the beam to higher energies is \dots the size of the~machine. Figure \ref{lhc_geo} at the very beginning of this paper gives a nice impression about that fact.

\subsection{Focusing Properties}
In addition to the main bending magnets that guide the beam on a closed orbit, focusing fields are needed to keep the particles close together. In modern storage rings and light sources the~particles are kept for many hours in the machine and a carefully designed focusing structure is needed to provide the necessary beam size at different locations in the ring and guarantee stability of the~motion.
\par
\begin{figure}[h!]
\begin{center}
\includegraphics[width=0.35\columnwidth] {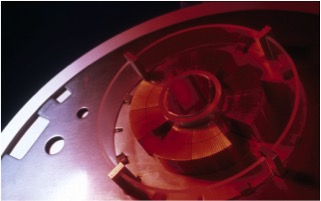}
\caption{Super conducting quadrupole of the LHC storage ring.%
}
\label{LHC_quad}
\end{center}
\end{figure}

 
Following the example of classical mechanics, linear restoring forces are needed, just as in the case of a harmonic pendulum. Quadrupole magnets provide the corresponding field:
They create a magnetic field that depends linearly on the particle amplitude, i.e. the distance of the~particle from the design orbit.
\begin{equation}
B_{x}=g\cdot y   \hspace{2cm}
B_{y}=g \cdot x
\end{equation}
The constant $g$ is called gradient of the magnetic field and characterises the focusing strength of the~magnetic lens in both transverse planes. For convenience it is - as the dipole field - normalised to the particle momentum. The normalised gradient is called $k$ and defined by 
\begin{equation}
k=\frac{g}{p/e}=\frac{g}{B \rho}.
\end{equation}
The technical layout of such a quadrupole is shown  in  Fig. \ref{LHC_quad}, that - again in the case of the LHC dipoles - shows a super conducting magnet.

Now, that we defined the basic building blocks of a storage ring, the task of the designer will be to arrange these in a so-called magnet lattice and to optimise the field strengths in a way to obtain the~required beam parameters. 
\par
\begin{figure}[h!]
\begin{center}
\includegraphics[angle = -90, width=0.7\columnwidth] {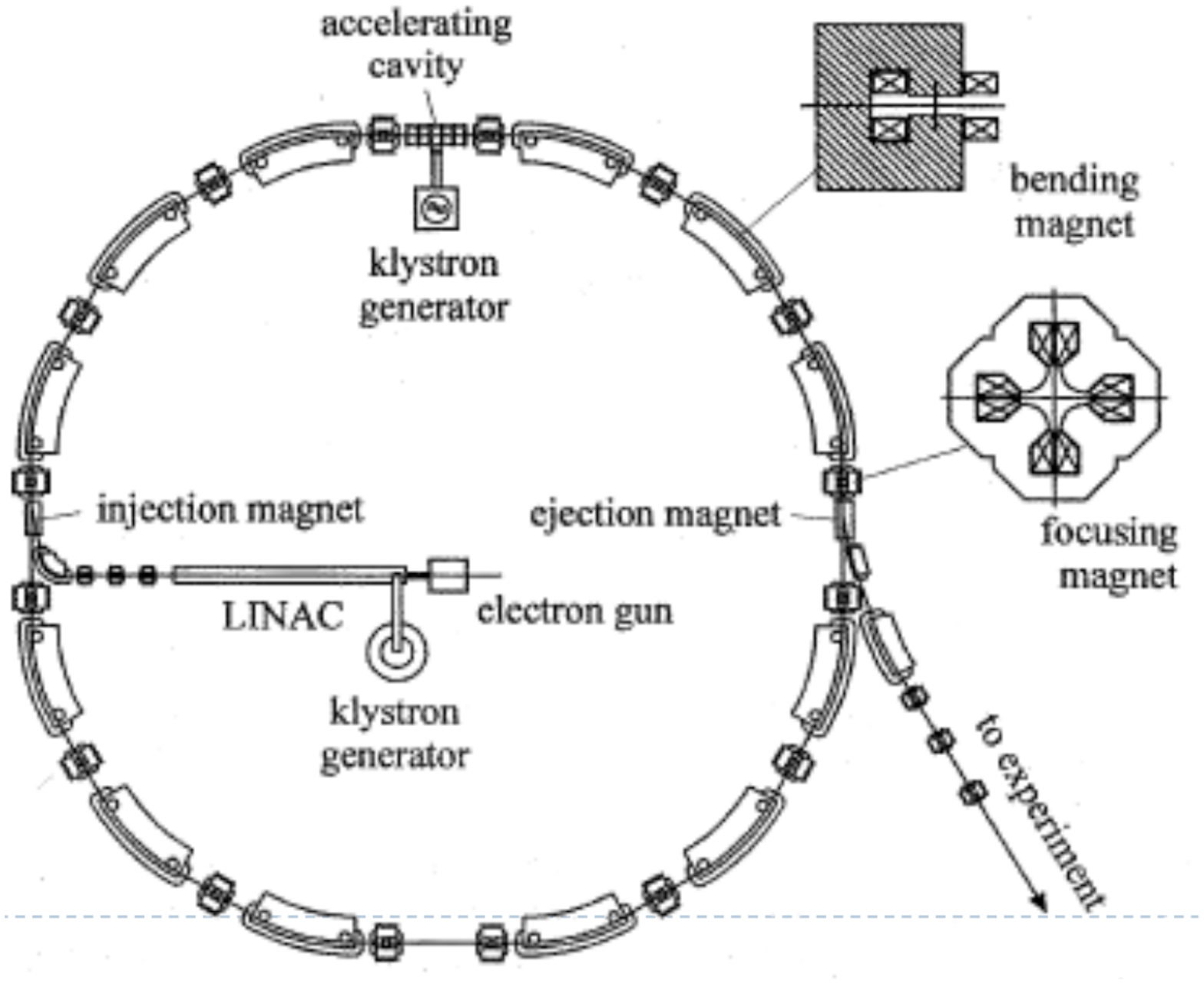}
\caption{Schematic layout of a synchrotron.}
\label{synchrotron_schematic}
\end{center}
\end{figure}

A general design principle of modern synchrotrons or storage rings should be pointed out here:
In general these machines are built following a so-called separate function scheme. Schematically this is shown in Fig. \ref{synchrotron_schematic}. 

\begin{figure}[h!]
\begin{center}
\includegraphics[width=0.5\columnwidth] {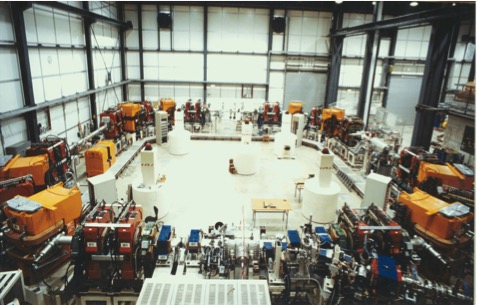}
\caption{TSR storage ring, Heidelberg, as a typical example of a separate function strong focusing storage ring.%
}
\label{TSR_photo}
\end{center}
\end{figure}

An example of how such a magnet lattice looks like in reality is given in  Fig. \ref{TSR_photo}. It shows the~dipole (orange) and quadrupole (red) magnets in the TSR storage ring in Heidelberg.  Eight dipoles are used to bend the beam on a "circle" and the quadrupole lenses in between them provide the focusing to keep the particles within the aperture limits of the vacuum chamber. 
Every magnet is designed and optimised for a certain task: bending, focusing, chromatic correction, etc. We separate the magnets in the design according to the job they are supposed to do and only in rare cases a combined function scheme is chosen, where different magnet properties are combined in one piece of hardware. Expressed mathematically, we refer to the general Taylor expansion of the~magnetic field: 
\begin{equation}
\frac{B(x)}{p/e}=\frac{1}{\rho}+ k \cdot x + \frac{1}{2 !}mx^{2} + \frac{1}{3 !}nx^{3}+ ...
\label{b_eq}
\end{equation}
Following the arguments above, we take at the moment only constant (dipole) or linear terms (quadrupole) in Eq. (\ref{b_eq}) into account. 
The higher order field contributions will be treated afterwards as - hopefully - small perturbations.

\begin{figure}[h!]
\begin{center}
\includegraphics[width=0.4\columnwidth] {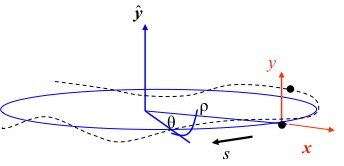}
\caption{Coordinate system used in particle beam dynamics. The longitudinal coordinate {\it s} is moving around the~ring with the particle considered.%
}
\label{Coordsystem}
\end{center}
\end{figure}

The particle will now follow the "circular path'', defined by the dipole fields, and in addition will perform harmonic oscillations in both transverse planes. Schematically the situation is shown in Fig. \ref{Coordsystem}. An ideal particle will follow the design orbit that is idealised in the plot by a circle. Any other particle will perform transverse oscillations under the influence of the external focusing fields  and the amplitude of these oscillations will finally define the beam size.

\begin{figure}[h!]
\begin{center}
\includegraphics[width=0.27999999999999997\columnwidth] {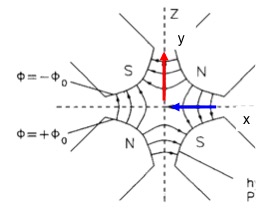}
\caption{Field configuration in a quadrupole magnet and the direction of the focusing and defocusing forces in both planes.%
}
\label{Quad_field}
\end{center}
\end{figure}

 Unlike a classical harmonic oscillator, however, the equation of motion in the horizontal and vertical plane differ a little bit:
Assuming a horizontal focusing magnet, the equation of motion is
\begin{equation}
x''+x \cdot (\frac {1}{\rho^{2}}+k)=0,
\end{equation}
where the derivative refers to the $s$-coordinate, $x'=dx/ds$ , which is nothing else than the angle of the~trajectory, and $x''=d^2x/ds^2$. k describes  the normalised gradient, as introduced above and the $1/ \rho^{2}$ term reflects the so-called weak focusing, which is a property of the bending magnets.
In the orthogonal, vertical plane however, due to the orientation of the field lines -- which clear enough results in the end from Maxwell's equations -- the forces turn into a defocusing effect. And -- also clear enough -- the weak focusing term disappeared, as usually the machine will be built in the horizontal plane and no vertical bends will be considered.
\begin{equation}
y''- y \cdot k=0
\end{equation}
The principle problem, arising from the different directions of the Lorentz force in the two transverse planes of a quadrupole field  is sketched  in Fig. \ref{Quad_field}. It is the task of the machine designer to find an~adequate solution to this problem and define a magnet pattern that will provide an overall focusing effect in both transverse planes.

Following quite closely the example of the classical harmonic oscillator,
we can write down the~solutions of the above mentioned equations of motion. We refer for simplicity to the situation in the~horizontal plane; a "focusing" magnet is thus focusing in this horizontal plane and at the~same time defocusing in the vertical one.
Starting with initial conditions for the particle amplitude $x_{0}$ and angle $x'_{0}$ in front of the magnet element we obtain the following relations for the trajectory inside the magnet:
\begin{equation}
x(s)=x_{0}\cdot cos(\sqrt{{|K|}}s)+x'_{0}\cdot \frac{1}{\sqrt{{|K|}}}sin(\sqrt{{|K|}}s)
\end{equation}

\begin{equation}
x'(s)=-x_{0}\cdot \sqrt{{|K|}} sin(\sqrt{{|K|}}s)+x'_{0}\cdot cos(\sqrt{{|K|}}s).
\end{equation}

Here - for convenience and following the usual treatment in literature - the strong and weak focusing effects, k and $1/\rho^2$, are merged in one parameter, $K=k+1/\rho^2$. Usually these two equations are combined in a more elegant and convenient matrix form.

\begin{equation}
\left(
\begin{array}{c}
x \\
x' 
\end{array} 
\right)_{s}
={\bf M_{\it{foc}}} \cdot 
\left(
\begin{array}{c}
x \\
x'
\end{array} 
\right )_{0},
\end{equation}
where the matrix $M_{foc}$ contains all relevant information about the magnet element:

$$
{\bf M_{{\it foc}}}=
\left(
\begin{array}{cc}
cos(\sqrt{{|K|}})s              &     \frac{1}{\sqrt{{|K|}}} sin(\sqrt{{|K|}})s  \\
- \sqrt{{|K|}}sin(\sqrt{{|K|}})s   &   cos(\sqrt{{|K|}})s  
\end{array} 
\right).
$$

\begin{figure}[h!]
\begin{center}
\includegraphics[width=0.35\columnwidth] {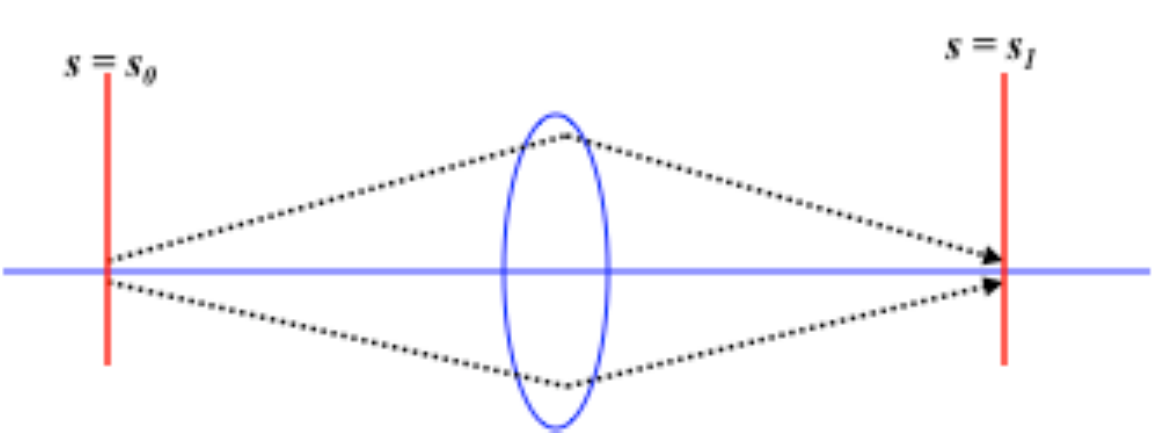}
\caption{Schematic principle of the effect of a focusing quadruole magnet.%
}
\label{Matrix_foc}
\end{center}
\end{figure}

Schematically the situation is visualised in Fig. \ref{Matrix_foc}. 

\begin{figure}[h!]
\begin{center}
\includegraphics[width=0.35\columnwidth] {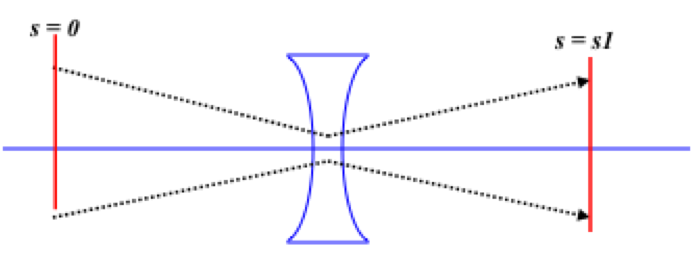}
\caption{Schematic principle of the effect of a de-focusing quadrupole magnet.%
}
\label{Matrix_defoc}
\end{center}
\end{figure}

In the case of a defocusing magnet we obtain  in full analogy, 
\begin{equation}
\left(
\begin{array}{c}
x \\
x' 
\end{array} 
\right)_{s}
={\bf M_{{\it defoc}}} \cdot 
\left(
\begin{array}{c}
x \\
x'
\end{array} 
\right )_{0}
\end{equation}

with

$$
{\bf M_{{\it defoc}}}=
\left(
\begin{array}{cc}
cosh(\sqrt{{|K|}}s)              &     \frac{1}{\sqrt{{|K|}}} sinh(\sqrt{{|K|}}s  \\
\sqrt{{|K|}}sinh(\sqrt{{|K|}})s   &   cosh(\sqrt{{|K|}})s  
\end{array} 
\right),
$$
see   Fig. \ref{Matrix_defoc}.    

For completeness we include finally the case of a field free drift of length $s$. Putting $K=0$ we obtain

$${\bf M_{{\it drift}}}=
\left(
\begin{array}{cc}
1          &     s  \\
0          &     1
\end{array} 
\right).
$$

This matrix formalism allows in a quite elegant way to combine the elements of a storage ring and calculate straight forward the particle trajectories. As an example we refer here to the~simple case of an alternating focusing and defocusing lattice, a so-called FODO lattice. Knowing the~properties of every single element in the accelerator, we can establish the corresponding matrices and calculate step by step the amplitude and angle of the single particle trajectory around the~ring.
Even more convenient and mathematically simple, we can multiply out the different matrices and -- given the initial conditions $x_{0}, x'_{0}$ -- get directly the trajectory at any location in the~ring,~e.g.,



\begin{equation}
{\bf M_{total}} =  {\bf M_{foc}} \cdot {\bf M_{drift}}\cdot {\bf M_{dipole}}\cdot {\bf M_{drift}} \cdot {\bf M_{defoc}}\ldots
\end{equation}

\begin{figure}[h!]
\begin{center}
\includegraphics[width=0.7\columnwidth] {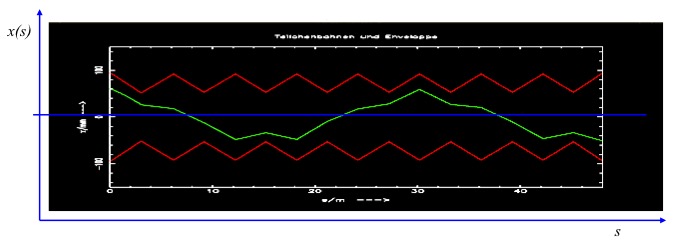}
\caption{Calculated particle trajectory in a simple storage ring.%
}
\label{track_1}
\end{center}
\end{figure}

Schematically the trajectory obtained is shown in Fig. \ref{track_1} .

Several facts have to be emphasised in this context:
\begin{itemize}
\item At each moment, or in each lattice element, the trajectory follows  a harmonic oscillation.
\item However due to the different restoring or defocusing forces, the solution will look different at each place.
\item In the linear approximation that we refer to in this context, each and every particle will feel the~same external fields and their trajectories differ only due to their different initial conditions.
\item There seems to be an overall oscillation in both transverse planes while the particle is travelling around the ring. Its amplitude stays well within the boundaries set by the vacuum chamber and its frequency, which in the example of  Fig. \ref{track_1} is roughly 1.4 transverse oscillations per revolution, corresponds to the eigen-frequency of the particle under the influence of the external fields.
\end{itemize}

\begin{figure}[h!]
\begin{center}
\includegraphics[width=0.42\columnwidth] {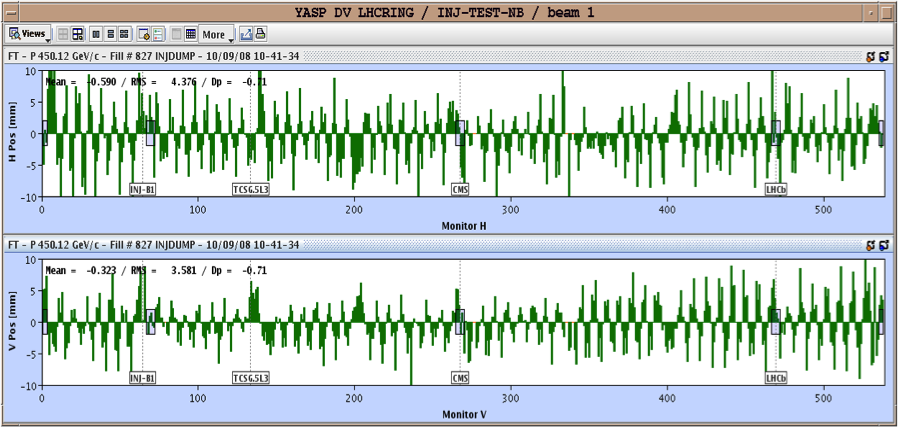}
\caption{Horizontal (top) and vertical (bottom) closed orbit oscillations, measured  in LHC during the commissioning of the machine.%
}
\label{LHC_orbit}
\end{center}
\end{figure}

Coming closer to a real existing machine, we show in Fig. \ref{LHC_orbit} the orbit that has been measured during one of the first injections into the LHC storage ring. In the upper part of the figure the horizontal, in the lower part the vertical orbit oscillations are plotted on a scale of $\pm 10$ mm. Every histogram bar indicates the value of a beam position monitor at a certain location in the ring: The~orbit oscillations are clearly visible. Counting (or better fitting) the number of oscillations in both transverse planes we obtain the following values of 
\begin{equation}
Q_{x}=64.31 \\
\end{equation}
\begin{equation}
Q_{y}=59.32.
\end{equation}
These values, which describe the eigen-frequencies of the particles, are called horizontal and vertical tune. Knowing the revolution frequency we easily can calculate the transverse oscillation frequencies, which for this type of machine usually lies in the range of kHz. 

\begin{figure}[h!]
\begin{center}
\includegraphics[width=0.42\columnwidth] {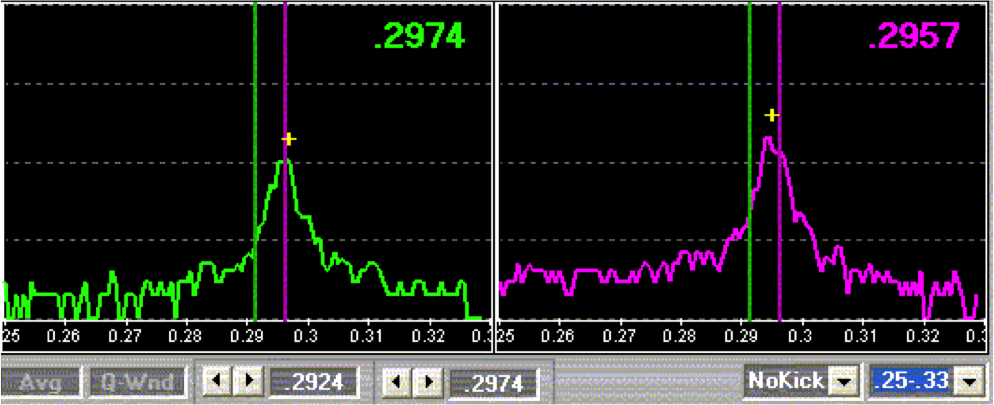}
\caption{Tune signal of a proton storage ring (HERA-p).%
}
\label{HERA_tune}
\end{center}
\end{figure}

As the tune characterises the particle oscillations under the influence of all external fields it is one of the most important parameters of the storage ring. Therefore, usually it is displayed and controlled at any time  in the control system of such a machine. Figure \ref{HERA_tune} shows as an example the tune signal of the~HERA proton ring \cite{HERA}. 
It is obtained via a Fourier analysis of the spectrum measured from the~signal of the complete particle ensemble. The peaks obtained indicate the two tunes in the horizontal and vertical plane of the machine and in a sufficiently linear machine a quite narrow spectrum is obtained.

\begin{figure}[h!]
\begin{center}
\includegraphics[width=0.48\columnwidth] {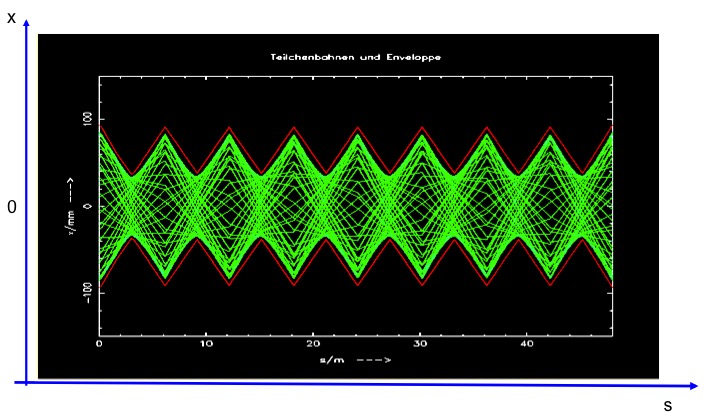}
\caption{Many single particle trajectories form in the end a pattern that corresponds to the beam size in the~ring.%
}
\label{track_n}
\end{center}
\end{figure}

Referring for a milli-second again to Fig. \ref{track_1}, the question arises, how the trajectory of the~particle will look like for the second turn, or for the third .... or for an arbitrary number of them. Now, as we are dealing with a circular machine the amplitude and angle, x, x', at the end of the first turn are the initial conditions of the second turn  and so on. And after many turns the overlapping trajectories begin to form a pattern, like in Fig. \ref{track_n},  that looks indeed like a beam having here and there a larger and smaller beam size but still being well defined in its amplitude by the external focusing forces. 
\par
And to make a long story short \cite{floquet}, a mathematical function, called $\beta$ or amplitude function, can be defined that describes the envelope of the single particle trajectories. 

\subsection{Limit IV: beam quality \dots also known as "emittance"}
Referring to this new variable, $\beta$, we can re-write the equation for a particle trajectory in its transverse oscillations as 
\begin{equation}
x(s)=\sqrt{\epsilon}\sqrt{\beta(s)}\cdot cos(\psi(s) + \phi),
\label{epsbeta}
\end{equation}
where $\psi$ describes the phase of the oscillation, $\phi$ its initial phase condition and $\epsilon$ is a characteristic parameter of the single particle, or considering a complete beam now, of the ensemble of particles. Indeed $\epsilon$ describes the space occupied by the particle in the transverse (here simplified two dimensional) phase space x, x'. And a bit more correct: The area in x, x' space that is covered turn by turn by the~particle's coordinates is given by
\begin{equation}
    A=\pi \cdot \epsilon
\end{equation}
and as long as we consider conservative forces acting on the particle this area according to Liouville's theorem is constant. Here we take these facts as given, but we would like to point out that as a direct consequence the so-called emittance $\epsilon$ cannot be influenced with what ever external fields. It is a property of the beam and we have to take it as given \dots and handle it with care. 
\par
\begin{figure}[h!]
\begin{center}
\includegraphics[width=0.35\columnwidth] {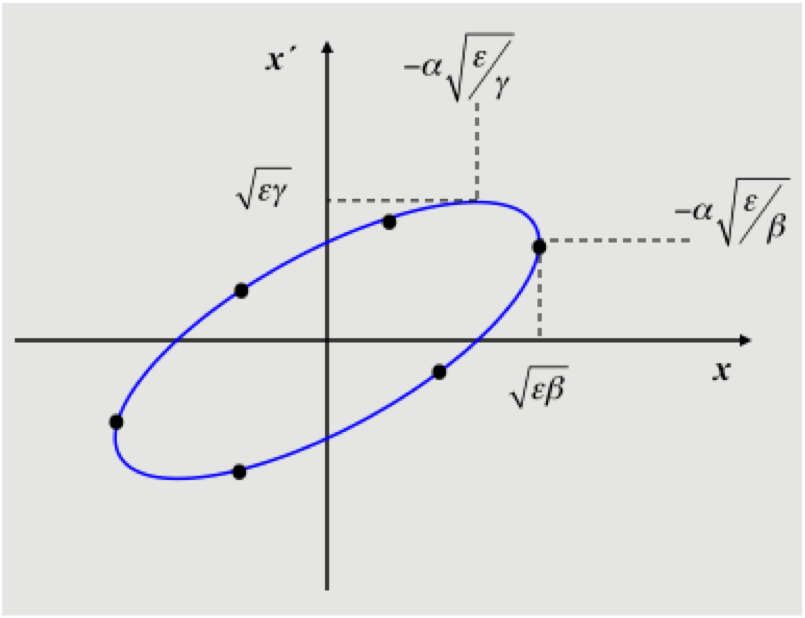}
\caption{Ellipse in x-x' phase space.}
\label{Ellipse}
\end{center}
\end{figure}

A bit more precise and following the usual treatments, we can draw the phase space ellipse of the~particles transverse motion, like for example illustrated  {Fig. \ref{Ellipse}}. The parameter $\alpha$ is used in text books to describe the derivative of the $\beta-$ function, $\alpha=-\frac{1}{2}\beta'$ and $\gamma = \frac{1+\alpha^2}{\beta}$. While the shape and orientation is defined by these optics functions, the area covered in phase space is constant -- as long as conservative forces are considered.
The concept of phase space is treated in much more detail in a later chapter of this book $\dots $  including the fact that often we prefer to call it {\it trace space} instead.

\begin{figure}[h!]
\begin{center}
\includegraphics[width=0.33\columnwidth] {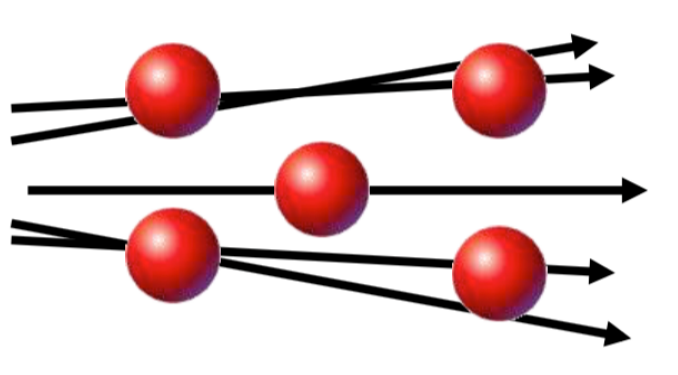}
\caption{Schematic picture of the trajectories in a beam. Small emittance means high quality of the particle ensemble, which in turn means small amplitudes and angles of the trajectories.
}
\label{emit_massimo}
\end{center}
\end{figure}

Talking a bit more about the beam as an ensemble of many (typically $10^{11}$) particles, and referring to Eq. (\ref{epsbeta}), at a given position in the ring the beam size is defined by the emittance $\varepsilon$ and the beta function $\beta$. Thus for a moment in time the cosine term in Eq. (\ref{epsbeta}) will be one and the trajectory amplitude will reach its maximum value. Now assuming that we consider a particle at one sigma of the transverse density distribution, and referring to the emittance of this reference particle we can calculate the size of the complete beam in the sense that the complete area (within one sigma) of all particles in (x, x') phase space is surrounded (and thus defined) by our one sigma candidate. Visualised a bit more qualitatively, we refer to Fig. \ref{emit_massimo}. A small emittance means  small amplitudes of the particles' trajectories and small divergence. The more parallel (Laser-like) and the more dense, the better; as easy as that.

\par
\begin{figure}[h!]
\begin{center}
\includegraphics[width=0.33\columnwidth] {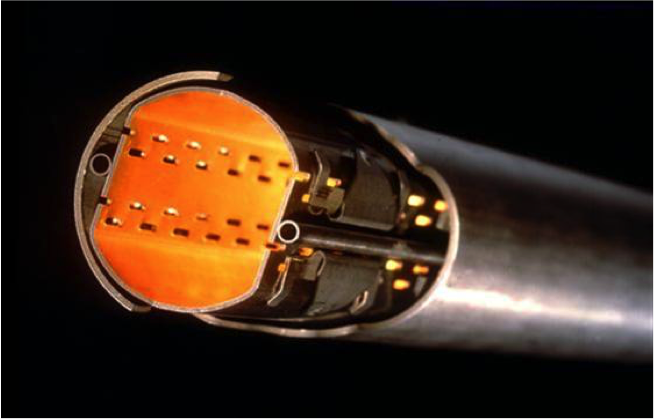}
\caption{The LHC vacuum chamber with the beam screen to shield the super conducting magnet bore from synchrotron radiation.%
}
\label{Beam_screen}
\end{center}
\end{figure}

In this sense the value $\sqrt{\epsilon \cdot \beta(s)}$ will define the one sigma beam size in the transverse plane. 
As an example the values for the LHC proton beam are used:
In the periodic pattern of the arc the beta function is $\beta=180$ m and the emittance at flat top energy is roughly $\varepsilon=5 \cdot 10^{-10}$ rad m.
The resulting typical beam size therefore is 0.3 mm. Now, clear enough, we will not design a~vacuum aperture of the machine based on one sigma beam size. Typically an aperture requirement corresponding to 12$\sigma$ is a good rule to guarantee sufficient aperture including tolerances from magnet misalignments, optics errors and operational flexibility. In Figure \ref{Beam_screen} the LHC vacuum chamber is shown, including the beam screen to protect the cold bore from synchrotron radiation. Its aperture radius corresponds to a minimum number of 18 $\sigma$ beam size.

\subsection {Limit V: The beam current}
\begin{figure}[h!]
\begin{center}
\includegraphics[width=0.5\columnwidth] {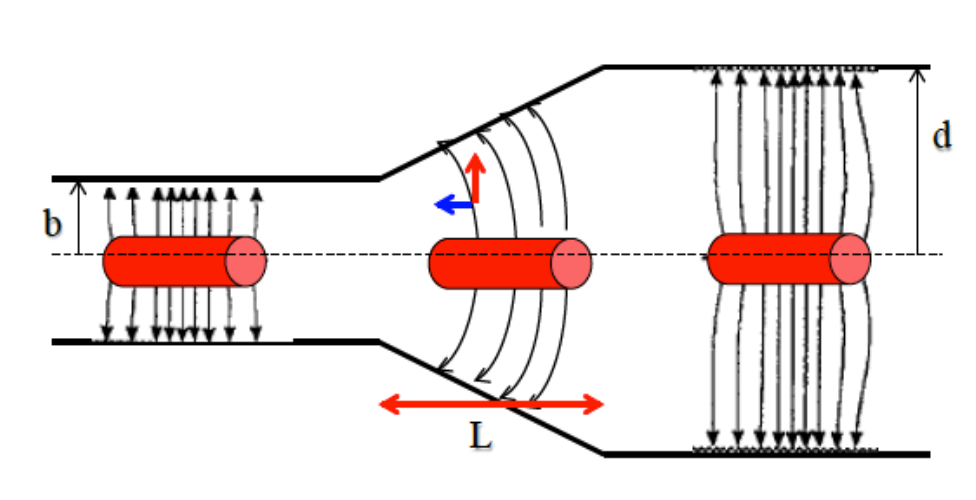}
\caption{Schematic view of the wake fields induced due to a sudden change of the vacuum chamber geometry.%
}
\label{wake_lines}
\end{center}
\end{figure}

Beyond the beam quality, expressed as "emittance" in the previous paragraph, the beam current, stored or accelerated in our machine is subject to limitations. Keywords here are wake fields, machine impedance, and single and multi-bunch instabilities related to that. The topic is worth a book of its own (see e.g. Ref. \cite{Massimo_wake}) and we only can mention a simple example here. 
Until now we have described the beam as an ensemble of individual particles, moving free through the~accelerator. This picture however is oversimplified. In reality the beam is a highly charged dollop  of particles and, as a matter of fact, talks via its electro-magnetic field to the environment. While this is usually metallic - first of all clearly the vacuum chamber - an image current is induced, running in parallel to our beam and creating at any moment funny shapes of field lines that act back on our primary beam.
The effect is described as "impedance" and the corresponding fields can severely distort the stability of the beam. Qualitatively, as described in Fig. \ref{wake_lines}, the effect is most serious whenever changes in the geometry of the environment occur. However, as soon as we accumulate more and more particles in our machine, we will sooner or later hit the limit of beam stability, created by wake fields.

For the simple case of the situation shown in Fig. \ref{wake_lines} and a round beam, the longitudinal field created leads to a power loss of the beam that depends on the step in the vacuum chamber geometry and on the square of the stored beam current.
\begin{equation}
P_b=\frac{I^2}{2 \pi \varepsilon_0 v} ln{\frac{d}{b}}
\end{equation}
where $I$ denotes the stored beam current, {\it v} the velocity of the bunch and {\it d} and { \it b} the vacuum chamber radius as indicated in the figure.

\section {Particle Colliders}
\subsection {Limit VI: Fixed Target Collider}

\begin{figure}[h!]
\begin{center}
\includegraphics[width=0.4\columnwidth] {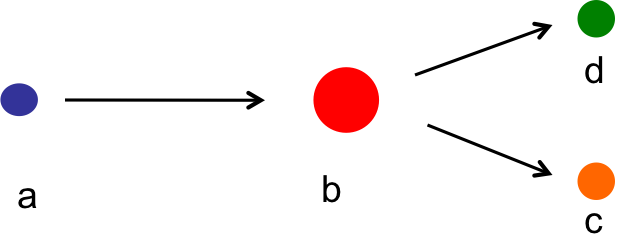}
\caption{Schematics of a moving particle $A$ colliding with a target particle $B$ at rest.%
}
\label{fixed_target_scheme}
\end{center}
\end{figure}

The easiest way to do physics using particle accelerators, is to bang the accelerated beam onto a~target and analyse the resulting events. 
While in high energy physics we nowadays do not apply this technique very often anymore it still plays an essential role in the regime of atomic and nuclear physics experiments.
The advantage is: It is quite simple, once the accelerator has been designed and built and the~produced particles are easily separated due to the kinematics of the reaction. Schematically the situation is shown in Fig.  \ref{fixed_target_scheme}. The particle "A" which is produced and accelerated in the machine is directed onto the~particle "B" which is at rest in the laboratory frame. The~produced particles are named "C" and "D" in the~example. 

While the set up of such a scheme is quite simple, it is worth studying a bit the available energy, in the centre of mass system.
The relativistic overall energy is given by 
\begin{equation}
E^{2}=p^{2}c^{2}+m_0^{2}c^{4},
\end{equation}
which is true for a single particle but equally valid for an ensemble of particles. Now,  most important, the rest energy of the particle ensemble  is constant (sometimes called {\it invariant mass of the~system}). \\
Considering now the system of the two particles colliding, we therefore can write
\begin{equation}
(E^{cm}_{a}+E^{cm}_{b})^{2}-(p^{cm}_{a}+p^{cm}_{b})^{2} c^{2}=(E^{lab}_{a}+E^{lab}_{b})^{2}-(p^{lab}_{a}+p^{lab}_{b})^{2} c^{2}.
\end{equation}
In the frame of the centre of mass system we get by definition 
\begin{equation}
p^{cm}_{a}+p^{cm}_{b}=0
\end{equation}
while in the laboratory frame where particle "B" is at rest we have simply
\begin{equation}
p^{lab}_{b}=0.
\end{equation}
The equation for the invariant mass therefore simplifies to 
\begin{equation}
W^{2}=(E^{cm}_{a}+E^{cm}_{b})^{2}=(E^{lab}_{a}+m_{b} \cdot c^{2})^{2}-(p^{lab}_{a}\cdot c)^{2}.
\end{equation}
And neglecting for convenience the rest masses of the two particles, we get the quite simple expression 
\begin{equation}
W \approx \sqrt{2E^{lab}_{a} \cdot m_{b}\cdot c^{2}}. 
\end{equation}
In other words the energy that is available in the centre of mass system depends on the square root of the~energy of particle "A", which is the energy provided by the particle accelerator. A quite unsatisfactory situation !

\begin{figure}[h!]
\begin{center}
\includegraphics[width=0.4\columnwidth] {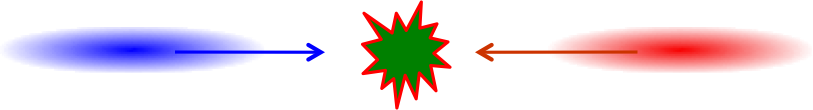}
\caption{Schematics of the collision of two colliding particle beams with equal energies.%
}
\label{coll_target_scheme}
\end{center}
\end{figure}

As a consequence, the design of modern high energy accelerators was naturally concentrated on the development of particle colliders, where two counter rotating beams are brought into collision at one or several interaction points (Fig. \ref{coll_target_scheme}).

\begin{figure}[h!] 
\begin{center}
\includegraphics[width=0.5599999999999999\columnwidth] {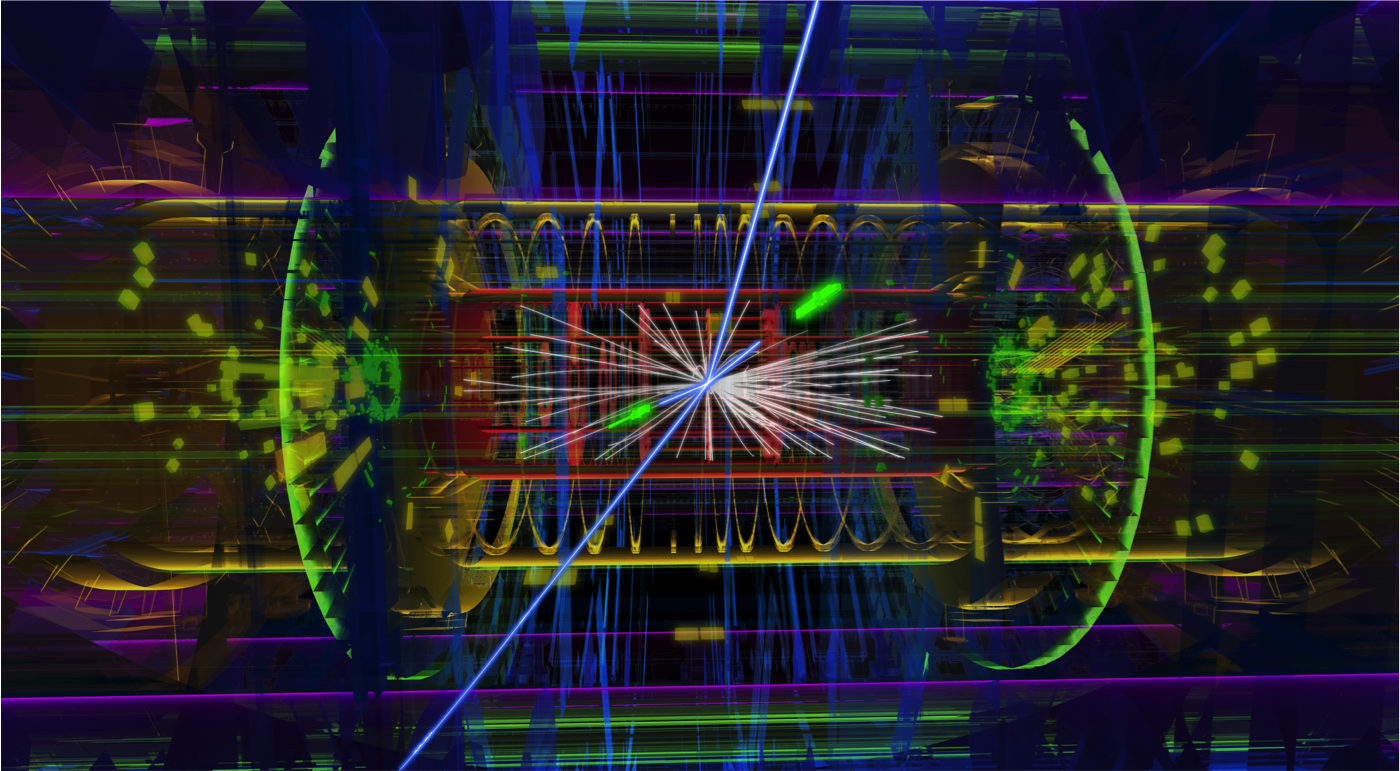}
\caption{"Typical" event observed in a collider ring: A Higgs particle in the ATLAS detector.  %
}
\label{Higgs_event}
\end{center}
\end{figure}

Calculating again, for the case of two colliding beams of equal particles and energies the~available energy in the centre of mass system, we get 
\begin{equation}
(p^{cm}_{a}+p^{cm}_{b})^{2}=0
\end{equation}
and by symmetry of the situation as well 
\begin{equation}
(p^{lab}_{a}+p^{lab}_{b})^{2}=0
\end{equation}
and so the full energy delivered to the particles in the accelerator is available during the collision process.
\begin{equation}
W= E^{lab}_{a}+E^{lab}_{b} =2\cdot E^{lab}_{a}. 
\end{equation}
A ``typical'' example of a high energy physics event in such a collider is shown in Fig. \ref{Higgs_event}.

\newpage
\subsection {Limit VII: The un-avoidable physics detectors}     
\begin{figure}[h!]
\begin{center}
\includegraphics[width=0.5599999999999999\columnwidth] {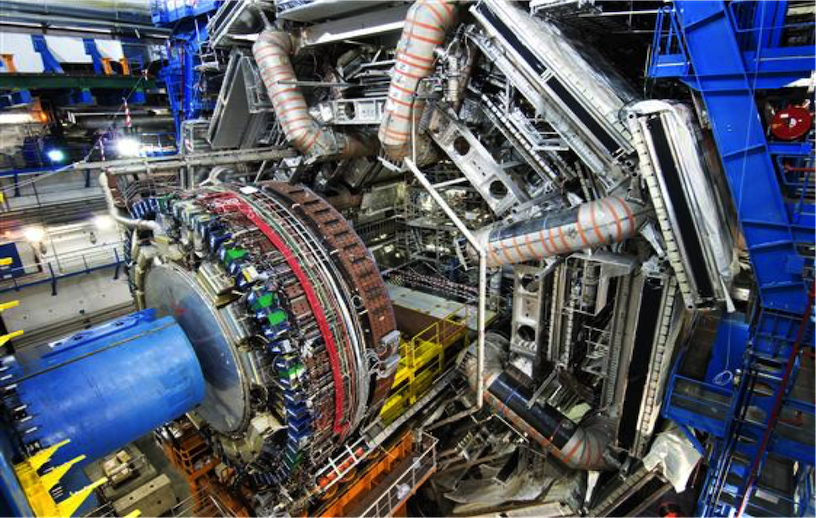}
\caption{ATLAS detector at LHC: 46 m in length, overall weight 7000 t.%
}
\label{Atlas}
\end{center}
\end{figure}

While it is quite clear that a particle collider ring is a magnificent machine in the quest for higher energies, there is a small problem involved, called ``particle detector''. In the arc of the storage ring usually we can find a nice pattern of magnets that provide us with a well defined beam size, expressed as beta-function. However special care has to be taken as soon as our colleagues from the high energy physics want to install a particle detector to analyse the events. Especially, working at the energy frontier, just like the~accelerators, these devices tend to grow considerably in size. 
In~{Figure~\ref{Atlas}} the largest particle detector installed in a storage ring is shown as impressive example: The ATLAS detector at LHC.

\par
\begin{figure}[h!]
\begin{center}
\includegraphics[width=0.5\columnwidth] {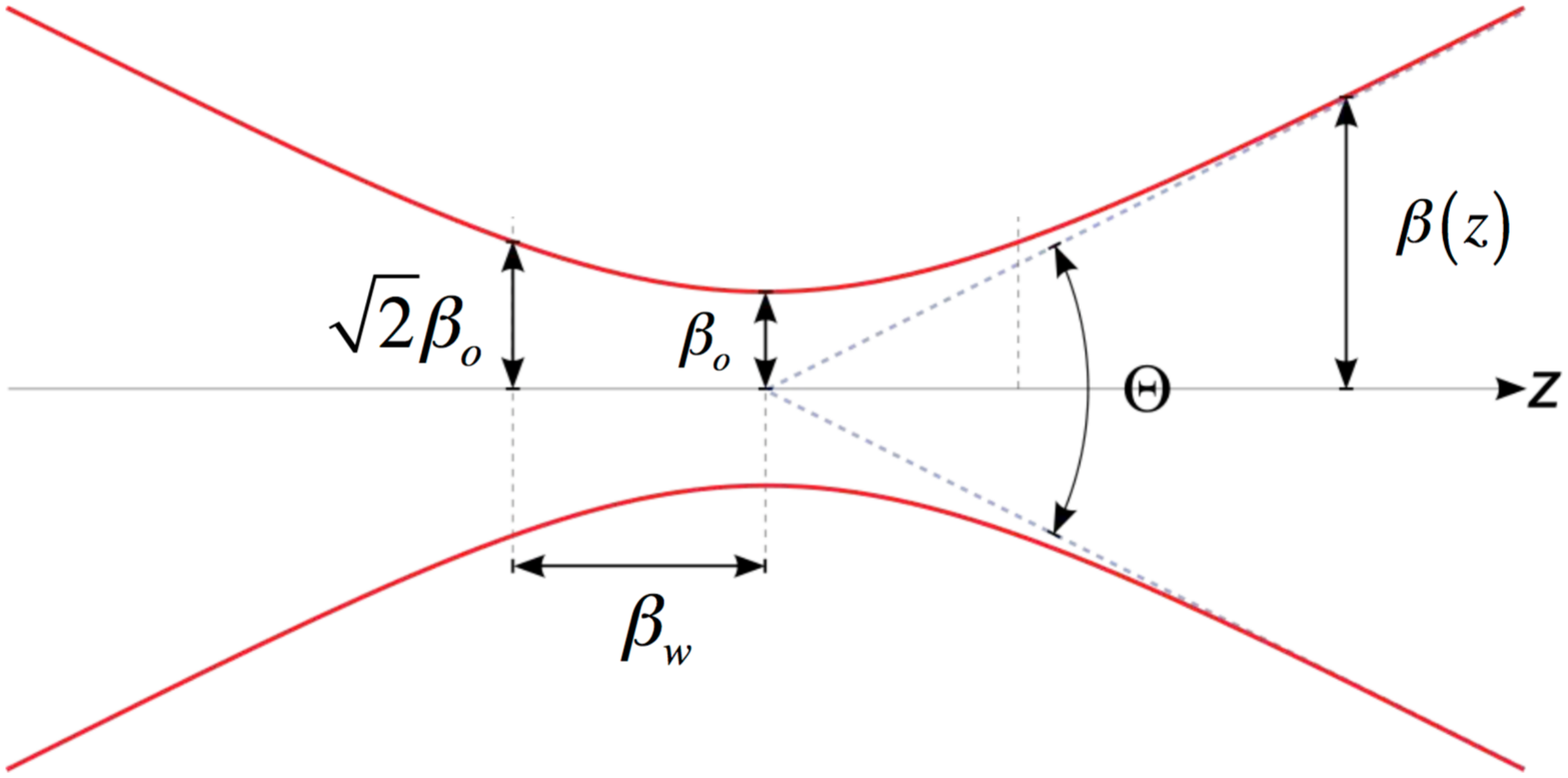}
\caption{The beam envelope in the neighborhood of a symmetric waist: the smaller the beta function at the~IP, the~faster the beam size is growing.}
\label{beam_envelope_IP}
\end{center}
\end{figure}

The design of the storage ring has to provide the space needed by the detector hardware and at the~same time create the smallest achievable beam spots at the collision point, which is usually right in the centre of the detector. Unfortunately, these requirements are a bit contradictory: The~equation for the~luminosity of a particle collider depends on the stored beam currents, $I_1, I_2$,  and the transverse spot size of the colliding beams at the Interaction Point (IP),  $\sigma_x^*,\sigma_y^* $:

\begin{equation} \label{lumi_eq}
L=\frac{1}{4 \pi e^{2}f_{0}b}\cdot \frac{I_{1}I_{2}}{\sigma_{x}^{*}\sigma_{y}^{*}}.
\end{equation}
The parameters $f_0$ and $b$ describe the revolution frequency of the ring and the number of stored bunches.    Unfortunately, however, the beta function in a symmetric drift grows quadratically as a~function of the~distance $s$ between the position of the beam waist and the first focusing element, namely:
\begin{equation}
\beta(s)=\beta_{0}+\frac {s^{2}}{\beta_{0}}.
\end{equation}
The smaller the beam size at the IP the faster it will grow until we can apply -- outside the detector region --  the first quadrupole lenses. For the beam size this translates into 
\begin{equation}
\sigma(s)=\sigma_0 \sqrt{1+\frac {(s-s_0)^{2}}{\beta_{0}}},
\end{equation}where $s_0$ refers to the position of the beam waist, e.g. in a collider the interaction point of the two beams. Schematically this fact is shown in Fig. \ref{beam_envelope_IP}. 

As a consequence this behaviour sets critical limits to the achievable quadrupole aperture - or, given the aperture, for the achievable  quadrupole gradient. Therefore the focusing lenses right before and after the IP, are placed as close as possible to the detector, and in general they are the~most critical and most expensive magnets in the machine: Their aperture need determines in the end the luminosity that can be delivered by the storage ring.
\par
For the experts we would like to add, that even if the bare aperture need can be fulfilled, another limit is usually reached:  the resulting chromaticity that is created in the mini-beta insertion and the~sextupole strengths that we will need to correct it,  usually present due to their non-linear fields serious limits for the stability of the particle motion; the resulting  so-called ``{\it dynamic aperture}''  is the next limit that we will face.

\subsection {Limit VIII: The relatively rareness of Nobel prize winning reactions}

The rate of physics events, generated in a particle collision process, does not only depend on the~characteristics of the colliding beams, but first of all, on the probability to create such an event, the so-called cross section of the process. 
\par
\begin{figure}[h!]
\begin{center}
\includegraphics[width=0.6\columnwidth] {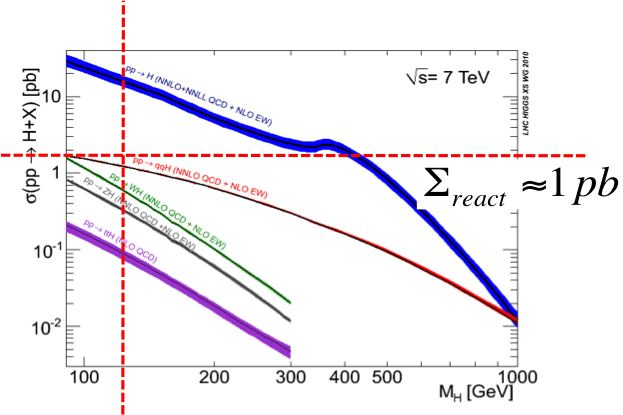}
\caption{Cross section of the Higgs for different production processes (court. CMS collaboration). %
}
\label{Higgs_cross}
\end{center}
\end{figure}

In the case of the Higgs particle, without doubt the high light of LHC Run1, the overall cross section is plotted in    Fig. \ref{Higgs_cross}. 

Without going into the details we can state that the cross section for a Higgs production is in the~order of 
\begin{equation}
\Sigma_{react} \simeq 1 pb.
\end{equation}
During the three years of the LHC Run1, namely in the years 2011-2013, an overall luminosity of 
\begin{equation}
\int Ldt = 25 fb^{-1} 
\end{equation}

has been accumulated. 

Combining these two numbers in the sense that the event rate of a reaction is given by $R=L \cdot \Sigma_{react}$ we get an overall number of produced Higgs particles 
of "some thousand". For a~Nobel prize winning event just at the edge of a reliable statistics.
As a consequence the particle colliders have to be optimised not only for highest achievable energies but at the same time for maximum stored beam currents and small spot sizes at the interaction points to maximise the~luminosity of the machine.

\subsection {Limit IX: The luminosity of a collider ring and the beam-beam effect}
Following the argumentation above, the design goal will be to prepare, accelerate and store two counter rotating particle beams to profit best from the energy of the two beams at the collision process. Still, there is a prize to pay: unlike fixed target experiments, where the "particle" density of the target material is extremely high, in the case of two colliding beams the event rate is basically determined by the transverse particle density that can be achieved at the interaction point. \\
Assuming Gaussian density distributions in both transverse planes the performance of such a collider is described by the luminosity formula  {Eq. (\ref{lumi_eq})}.

While the revolution frequency $f_{0}$ and the number of bunches per beam $b$ are in the end determined by the size of the machine, the stored beam currents $I_{1} $ and $I_{2}$ and the beam size at the~interaction point $\sigma_{x}^{*}, \sigma_{y}^{*}$ have their own limitations. 
\par 
The most serious one is the beam-beam interaction itself. During the collision process the~individual particles of counter rotating  bunches feel the electro-magnetic field of the opposing bunch. In the~case of a proton-proton collider this strong fields act like a defocusing lens, and have a strong impact on the tune of the bunches \cite{Herr}. 

\begin{figure}[h!]
\begin{center}
\includegraphics[width=0.33\columnwidth] {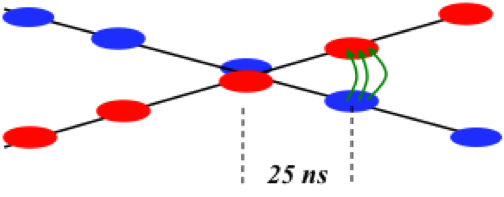}
\caption{Schematic view of the beam-beam interaction during the crossing of bunch trains.%
}
\label{bb_trains}
\end{center}
\end{figure}

\begin{figure}[h!]
\begin{center}
\includegraphics[width=0.55\columnwidth] {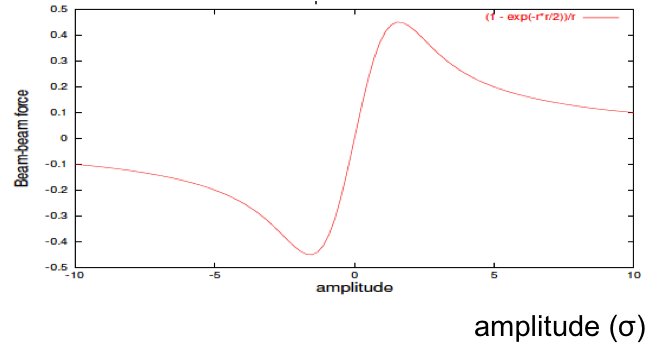}
\caption{Beam-beam force as a function of the transverse distance of the particle to the centre of the opposing bunch.%
}
\label{bb_force}
\end{center}
\end{figure}

In Figure \ref{bb_trains} the situation is shown schematically. Two bunch trains collide at the IP and during the collision process a direct beam-beam effect is observed. Even more, in addition to that, before and after the actual collisions, long range forces exist between the bunches that have a non-linear component as illustrated in Fig. \ref{bb_force}. As a consequence \cite{Herr} the tune of the beams is not only shifted with respect to the natural tune of the machine, but spread out, as different particles inside the bunches see different contributions from the beam beam interaction.

\begin{figure}[h!]
\begin{center}
\includegraphics[width=0.55\columnwidth] {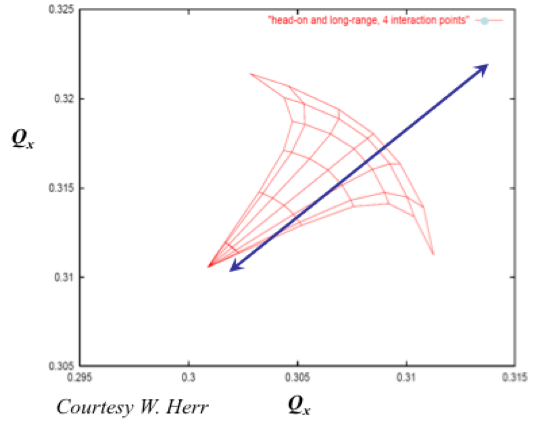}
\caption{Calculated tune shift due to the beam-beam interaction in LHC.%
}
\label{bb_working_diagram}
\end{center}
\end{figure}

In the tune diagram, which shows the horizontal and vertical tune in a common plot, therefore, we obtain not a single spot anymore, representing the ensemble of the particles, but a large array, that depends in shape, size and orientation on the particle densities, the distance of the bunches  at the long-range encounters and on the single bunch intensities. The effect has been calculated for the LHC and is shown in  Fig. \ref{bb_working_diagram}.

In linear approximation, i.e. referring to a distance of about 1-2 $\sigma$ the beam-beam force in Fig. \ref{bb_force} can be linearised and acts like a defocusing quadrupole effect. Accordingly a tune shift can be calculated to characterise the strength of the beam-beam effect in a collider. 
Given the parameters described above and introducing the classical particle radius $r_{p}$, the amplitude function at the~interaction point $\beta^{*}$ and the~Lorentz factor $\gamma$, we can express the tune shift due to the linearised beam-beam effect by 
\begin{equation}
\Delta Q_{y}=\frac{\beta_{y}^{*}\cdot r_{p} \cdot N_{p}}{2 \pi \cdot \gamma (\sigma_{x}+\sigma_{y})\sigma_{y}}.
\end{equation}
In the case of LHC, with a number of particles per bunch of $N_p=1.2*10^{11}$,  the design value of the~beam-beam tune shift is $\Delta Q=  0.0033 $ and in daily operation the machine is optimised to run close to this value, which puts the ultimate limit for the achievable bunch intensities in the collider.  
\newpage
\section {Electron Synchrotrons, Lepton Colliders and Synchrotron Light Sources}
\subsection {Limit X: Synchrotron-Light, the draw back of electron storage rings}
\begin{figure}[h!]
\begin{center}
\includegraphics[width=0.45\columnwidth] {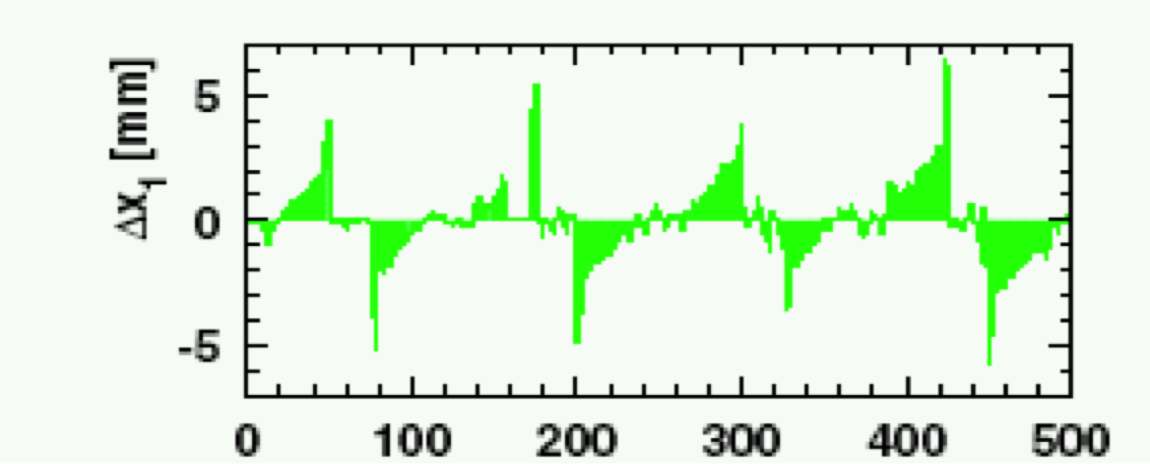}
\caption{Measured horizontal orbit of the LEP electron beam. Due to synchrotron radiation losses the~particle orbit is shifted towards the inner side of the ring in each arc.%
}
\label{saw_tooth}
\end{center}
\end{figure}

While in proton or heavy ion storage rings the design can follow more or less the rules that have been discussed above, 
the case changes drastically as soon as the particles gain more and more energy. Bent on a circular path, especially electrons will emit an intense light, so-called synchrotron radiation, that will have a strong influence on the beam parameters as well as on the design of the~machine. 
\\
Summarising here only briefly the situation, the power loss due to this radiation depends on the~bending radius and the energy of the particle beam:
\begin{equation}
P_{s}=\frac{2}{3}\alpha \hbar c^{2} \frac{\gamma^{4}}{\rho^{2}},
\label{syli}
\end{equation}
where $\alpha$ describes the fine structure constant, $\rho$ the bending radius in the dipole magnets of the~ring and $\gamma$ is the relativistic Lorentz factor. 
As a consequence the particles will lose energy turn by turn.
To compensate these losses, RF power has to be supplied to the beam at any moment. An example that visualises nicely the problem is shown in Fig. \ref{saw_tooth}. It shows the horizontal orbit of the former LEP storage ring. The electrons, traveling from  right to left in the plot lose a considerable amount of energy in each arc and as a consequence are deviating from the ideal orbit towards the inner side of the ring. The effect on the orbit is large: Up to 5 mm orbit deviations are observed. In order to compensate for these losses four RF stations are installed in the straight sections of the ring to provide the necessary power. 

The strong dependence of the synchrotron light losses on the relativistic $\gamma$ factor sets severe limits to the beam energy that can be carried in a storage ring of given size. And pushing for even higher energies means either the design of storage rings that are even  larger than the LEP or, in order to avoid synchrotron radiation, the design of linear accelerating structures. To make it even more clear: For a~given maximum power that can be re-supplied to the beam (and paid by the~laboratory), a factor of two increase of the energy of the stored electrons , means an increase of a factor of four in the size of the storage ring to be build.
\par
\begin{figure}[h!]
\begin{center}
\includegraphics[width=0.65\columnwidth] {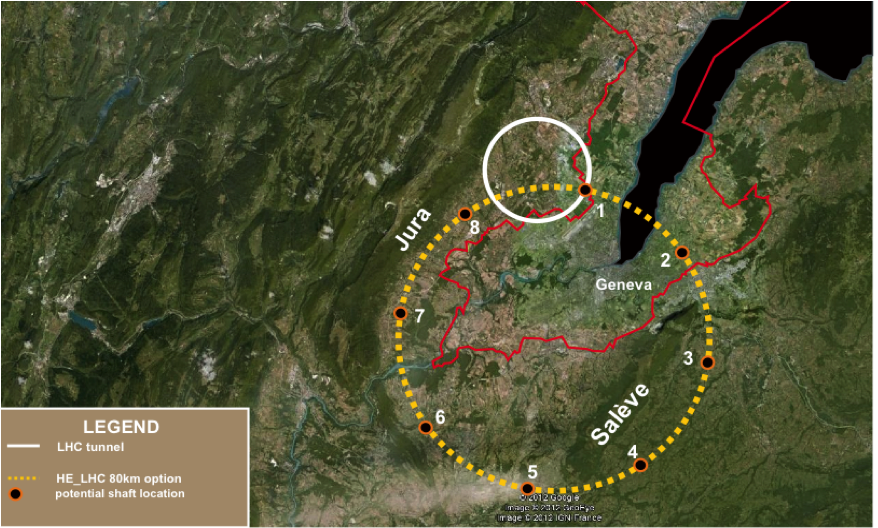}
\caption{Schematic view of a 100 $\! $ km long ring design in the Geneva region for the FCC study. %
}
\label{FCC}
\end{center}
\end{figure}

At present the next generation particle colliders are being studied \cite{TLEP} and the ring design of this Future Circular Collider (FCC) foresees a 100 km ring to carry electrons (and positrons) of up to 175 GeV beam energy. The dimension of this storage ring is far beyond of what has been designed up to now. The impressive sketch of the machine layout is shown in Fig. \ref{FCC} where the~dashed circle refers to the foreseen geometry of the 100 km ring and the solid one represents the little LHC machine.

For the maximum foreseen electron energy of E = 175 GeV, the synchrotron radiation would lead to an energy loss of 8.6 GeV per turn or an overall power of the radiated light of 47 MW at full beam intensity. Along with the power loss of the beam due to radiation goes a severe impact on the beam emittance. In hadron machines (in general, in beams where the energy loss due to synchrotron radiation is negligible) the beam emittance as defined above can be considered as constant for a given energy. Even better, if accelerated, it will shrink inversely as a function of the~energy:
\begin{equation}
\varepsilon \propto \frac{1}{\gamma} \hspace{0.5cm}  (hadrons).
\end{equation}
In lepton rings just the opposite is true:
Under the influence of the emitted radiation, which is a~quantum effect, an equilibrium is obtained between radiation damping and excitation and we get 
\begin{equation}
\varepsilon = \frac{55}{32 \sqrt{3}} \frac{\hbar c}{m_e c^2} \frac{\gamma^2}{J_x \rho} \frac{1}{\beta} \{{D^2+(\beta( D'+\alpha D)^2}\} \hspace{0.5cm}  (electrons).
\end{equation}
A kind of formidable equation; however if we focus on the essential issue here - namely the energy dependency -  we can simply point out that the emittance of a lepton beam in a ring is increasing quadratic as a function of the energy, $ \varepsilon \propto \gamma^2$ which makes it more and more difficult to achieve high luminosity or high brightness of the emitted light in an electron ring if we get to higher energies.

\subsection {Limit XXL: acceleration gradients in linear structures}
As far as lepton beams are concerned, ring colliders suffer from the severe limitation of synchrotron radiation and at a certain moment the construction of such a large facility does not seem reasonable anymore.
In order to avoid the synchrotron radiation, therefore, linear structures that had been discussed above and used in the infancy of particle accelerators are en vogue again. 
Still the~advantage of circular colliders cannot completely be neglected:
Even with a modest acceleration gradient in the RF structures the particles will get turn by turn a certain push in energy and they will reach some when the desired flat top energy in the ring. 
\par
\begin{figure}[h!]
\begin{center}
\includegraphics[width=0.7\columnwidth] {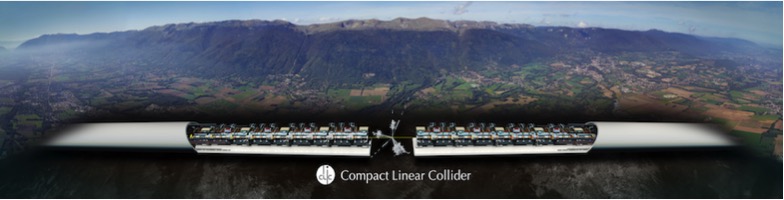}
\caption{Proposed location of the CLIC linear collider along the Jura mountain in Geneva region.%
}
\label{CLIC_photo}
\end{center}
\end{figure}

In a linear accelerator this repetitive acceleration is by design not possible. Within a single pass through the machine the particles will have to be accelerated to full energy. In order to keep the structure compact highest acceleration gradients therefore will be needed. One of the most prominent examples, proposed for a possible future collider is the CLIC design \cite{CLIC}. Within one passage through the 25 km long accelerator the electron  beam will get up to 3 TeV; and the same holds for the opposing positron beam. 
An artist view of tis machine is shown in Fig. \ref{CLIC_photo}.

\pagebreak
The CLIC main parameters are listed in Table {\ref{clic_param}}. Especially the accelerating gradient, i.e., the~energy gain per meter has to be emphasised. It has been pushed to the technically feasible maximum gradient, and the limit in the end is given by the breakdown of the electrical field in the~accelerating structure. 

\begin{table}[t]
\caption{The main parameters of the CLIC study.}
  \centering
  \begin{tabular}{|l|c|c|}
    \hline \hline
  		                                   &  \bfseries  500 GeV                     & \bfseries 3 TeV                          \\ 
		                                   \hline
Site Length                                 & 13 km                             &  48 km                         \\
Loaded accel. gradient [MV/m]  &                     \multicolumn{2}{c|} {12}                        \\ 
Beam power/beam [MW]            & 4.9                                 &  14                                \\
Bunch charge [$10^{9} $ e+/e]   & 6.8                                 &  3.7                                \\
Hor./vert. norm. emitt. [$10^{-6}/10^{-9}m$]     &       2.4/25   &  0.66/20                        \\ 
Beta Function [mm]                      &    \multicolumn{2}{c|} {10/0.07}                              \\
Beam Size at IP hor/vert [nm]      &   \multicolumn{2}{c|} {45/1}                                     \\
 \hline
 Luminosity ($cm^{-2} s^{-1} $      &  $2.3 \cdot 10^{34} $      &  $5.9 \cdot 10^{34} $  \\
 \hline \hline
  \end{tabular}
\label{clic_param}
\end{table}

\begin{figure}[tb]
\begin{center}
\includegraphics[width=0.7\columnwidth] {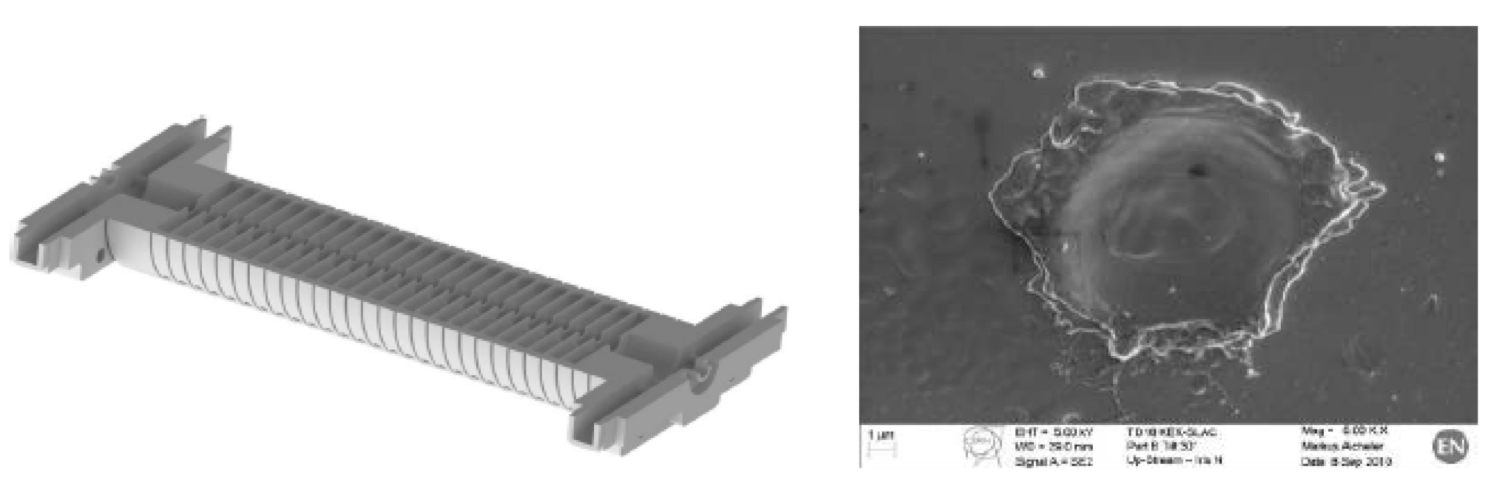}
\caption{Accelerating structure of the CLIC test facility CTF3; on the right side a electron microscope photo shows damage effects on the surface, created due to discharges in the module.%
}
\label{CLIC_structure}
\end{center}
\end{figure}

A picture of such a CLIC type structure is shown in Fig. \ref{CLIC_structure}. On the right hand side a~photo, taken with an electron microscope of the surface after a voltage breakdown is shown. At the~spot of the~sparking a little crater shows the possible damage of the surface and as a consequence the~deterioration of the~achievable gradient, which has to be avoided under all circumstances \cite{Palaia}. 

Being considerably higher than the typical values in circular machines, the gradient of $E_{acc}=100$ MV/m in a linear machine still leads to a design of bout 50 km overall length for a maximum achievable energy of $E_{max}=3$ TeV. 

\section {Conclusion}
\begin{figure}[tb]
\begin{center}
\includegraphics[width=0.6\columnwidth] {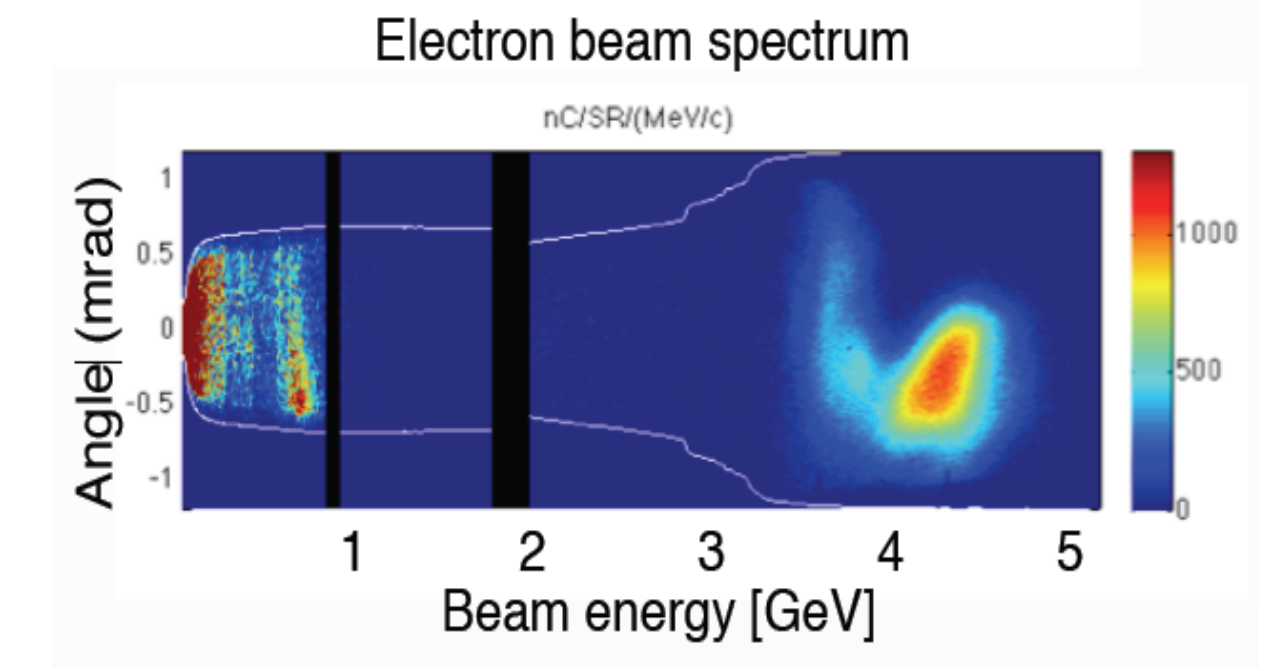}
\caption{electron beam accelerated in the wake potential of a plasma cell. Up to 4 GeV are obtained within a few cm length only \cite{pwa}.%
}
\label{pwa}
\end{center}
\end{figure}

Summarising the facts we state, that as soon as we talk about future lepton ring colliders (or to be more precise electron-positron colliders) the synchrotron losses set a severe limit to the achievable beam energy. And very soon the size of the machine will be beyond economical limits: For a~constant given synchrotron radiation loss the dimensions of the machine have to grow  quadratic with the beam energy. Linear colliders therefore are the proposed way to go. And in this case the~maximum achievable acceleration gradient is the key issue. New acceleration techniques, namely plasma based concepts where gradients are observed that are much higher than the present conventional techniques, are a most promising concept for these future colliders. An impressive example is shown in Fig. \ref{pwa}. Within a plasma cell of only some cm in  length, electrons are accelerated to several GeV. The gradients achievable are by orders of magnitude higher than in any conventional machine (see e.g., Ref. \cite{pwa}). Still there are problems to solve, like overall efficiency, beam quality (mainly the energy spread of the beam) and the achievable repetition rate. But we are convinced that it is worth studying this promising field.
And this is what this school was organised for.

\bibliography{bibliography/converted_to_latex.bib%
}


%

\end{document}